\documentclass[12pt,preprint]{aastex}
\usepackage{color}
\shorttitle{Time-Dependent models for Blazar Emissions}
\shortauthors{Asano et al.}

\begin{document}

\title{
Time-Dependent Models for Blazar Emissions with the Second-Order Fermi Acceleration
}
\author{\scshape Katsuaki Asano\altaffilmark{1}, Fumio Takahara\altaffilmark{2},
Masaaki Kusunose\altaffilmark{3}, Kenji Toma\altaffilmark{2}, and
Jun Kakuwa\altaffilmark{4}}
\email{asanok@icrr.u-tokyo.ac.jp}

\altaffiltext{1}{Institute for Cosmic Ray Research, The University of Tokyo,
5-1-5 Kashiwanoha, Kashiwa, Chiba 277-8582, Japan}
\altaffiltext{2}{Department of Earth and Space Science,
Osaka University, Osaka, 560-0043, Japan}
\altaffiltext{3}{Department of Physics, School of Science and Technology,
Kwansei Gakuin University, Sanda 669-1337, Japan}
\altaffiltext{4}{Department of Physical Science, Hiroshima University,
Higashi-Hiroshima, 739-8526, Japan}

\date{Submitted; accepted}

\begin{abstract}

The second-order Fermi acceleration (Fermi-II) driven by turbulence
may be responsible for the electron acceleration in blazar jets.
We test this model with time-dependent simulations.
The hard electron spectrum predicted by the Fermi-II process
agrees with the hard photon spectrum of 1ES 1101-232.
For other blazars that show softer spectra,
the Fermi-II model requires radial evolution
of the electron injection rate and/or diffusion coefficient
in the outflow.
Such evolutions can yield a curved electron spectrum, which
can reproduce the synchrotron spectrum of Mrk 421 from the radio to the X-ray regime.
The photon spectrum in the GeV energy range of Mrk 421 is hard to fit with a
synchrotron self-Compton model.
However, if we introduce an external radio photon field
with a luminosity of $4.9 \times 10^{38}~\mbox{erg}~\mbox{s}^{-1}$,
GeV photons are successfully produced via inverse Compton scattering.
The temporal variability of the diffusion coefficient or injection rate
causes flare emission.
The observed synchronicity of X-ray
and TeV flares implies a decrease of the magnetic field
in the flaring source region.

\end{abstract}


\maketitle

\section{Introduction}
\label{sec:intro}

Multi-frequency spectra of blazars are characterized by
the double peaks of the synchrotron and inverse Compton (IC) components.
They have been successfully fitted with steady-state
leptonic models \citep[e.g.][]{kin02,cel08}.
In most models, non-thermal emission is presumed
to be emitted by shock-accelerated electrons
\citep[the Fermi-I process; e.g.][]{kir98,spa01}.
The flare phenomena may be caused by internal shocks in the blazar outflows
as have been discussed in the models of the prompt emission of gamma-ray bursts
\citep[GRBs;][]{mes06}.

However, the emission from blazars, especially in quiescent states,
can be regarded as quasi-steady, which is different than that of GRBs.
The existence of steady shocks in the outflows is non-trivial.
This may imply a different acceleration process from the Fermi-I process.
The electron energy distributions obtained from the photon-spectrum fits also
cast doubt on the Fermi-I acceleration.
The maximum electron energy is far below the Bohm limit \citep{ino96},
while electrons accelerated by the shocks of supernova remnants
attain energies close to the Bohm limit \citep{aha99,yam04}.
The detections of very high-energy gamma-rays ($> 10^{11}$ eV)
from high-redshift blazars
\citep{1101EBL,aha07,1101}, despite obligatory absorption due to
extragalactic background light (EBL), indicate very hard photon spectra
(photon index $\lesssim 1.5$).
Those unusually hard spectra are supported by the non-detection
of GeV photons with {\it Fermi} \citep{ner10}.
The implied electron spectra may be harder
than the prediction of the simplest version of diffusive shock acceleration theory;
the electron spectral index should be larger than $2$.
Several mechanisms to produce harder spectra
for the shock accelerated particles have been proposed,
although they are not yet well established.
The non-linear back reaction of cosmic-ray pressure on the shock
structure \citep{mal01} has been frequently discussed.
Alternatively, \citet{vai99} considered the particle acceleration
in non-relativistic shocks and showed that the electron power-law spectral
indices can be smaller than 2 when the scattering center compression ratio
is larger than the gas compression ratio.
\citet{vai03} also demonstrated this for relativistic shocks.

Second order Fermi acceleration (Fermi-II) is a promising
process to make hard spectra \citep[e.g.][]{sch84,par95,bec06,sta08}.
This slow acceleration process can naturally explain
the lower maximum energy of electrons.
Applications of the Fermi-II to active galactic nucleus (AGN)
jet emission have been
discussed by several authors \citep[e.g.][]{bot99,sch00,kat06}.
The turbulence responsible for the Fermi-II acceleration
may be induced by the Kelvin--Helmholtz instability \citep{har04,miz07}
or the current-driven instability \citep{lyu99,nar09,miz11}.
Such instabilities may be triggered by recollimation of the jet
induced by a pressure gradient in the medium \citep{dal88,kom97,agu01}.
Actually, the signature of the Fermi-IIprocess has been explored in photon spectra.
X-ray spectra have been fitted with
a curved function, such as a log-parabolic shape,
which has been discussed in the theoretical context of
the Fermi-II process \citep{mas04,mas-mkn501,mrk421swift}.
Even for GRBs, the Fermi-II process has been considered \citep{asa09}
to yield hard spectra below the spectral peak energy ($\sim 0.1$--1 MeV).

\citet{lef11} adopted the Fermi-II process to fit a hard blazar
1ES 0229+200, although their discussion focused on the balance
between the acceleration and cooling.
In this paper, we further pursue the possibility of the Fermi-II process
in blazar jets.
Here we use the time-dependent code of \citet{asa11}
developed for GRB studies \citep[see also][]{asa12}
to follow the evolution of the electron energy distribution
and photon production.
In our code, the electron distribution is
obtained with the effects of the injection,
acceleration, radiative cooling, adiabatic cooling,
electron--positron pair production, and heating due to synchrotron self-absorption.
Based on the photon production and escape from the source region,
the code outputs photon spectra and lightcurves for an observer
including the Doppler and curvature effects.

We try to fit the broadband spectra
of 1ES 1101-232 and Mrk 421 with our simulations.
This would be the first application of comprehensive Fermi-II models
to current data of broadband blazar spectra
with a time-dependent method.\footnote{
The main purpose of \citet{lef11} is not to fit data,
while they fit the TeV spectrum of 1ES 0229+200
with a peculiar model, a Maxwell-like electron distribution
due to the balance between the acceleration and the cooling.
The time-dependent effects discussed there are not applied to the spectral fit.}
The temporal evolution of the electron and photon energy distributions
will be explicitly shown, which will help us understand the roles
of the temporal evolution of the Fermi-II process and
particle injection rates on the photon spectra.
We will show that the temporal evolution is important
not only for the spectral variability in flares \citep{kus00,time-dep}
but also for steady emission.
Temporal evolution of the injection rate and the acceleration efficiency etc.
may play an important role in steady photon spectra \citep[see e.g.,][]{bec06}.

%\begin{description}
% \item[Time-dependent code]\mbox{}\\
%\citet{time-dep}, ours are \citet{asa11,asa12}.
%\end{description}

In \S 2, we explain our model and numerical method.
The results for the hard spectrum blazar 1ES 1101-232
are shown in \S 3.
The results for the famous blazar Mrk421 are divided
into two parts: \S 4 for the steady photon spectrum
and \S 5 for the spectral variability in flares.
The summary and discussion are in \S 6.
To provide spectra and lightcurves for an observer,
we adopt the cosmological parameters $H_0=70~\mbox{km}~\mbox{s}^{-1}~\mbox{Mpc}^{-1}$,
$\Omega=0.3$, and $\Lambda=0.7$.

\section{Numerical Methods}
\label{sec:model}

\begin{figure}[htb!]
\centering
\epsscale{1.0}
\plotone{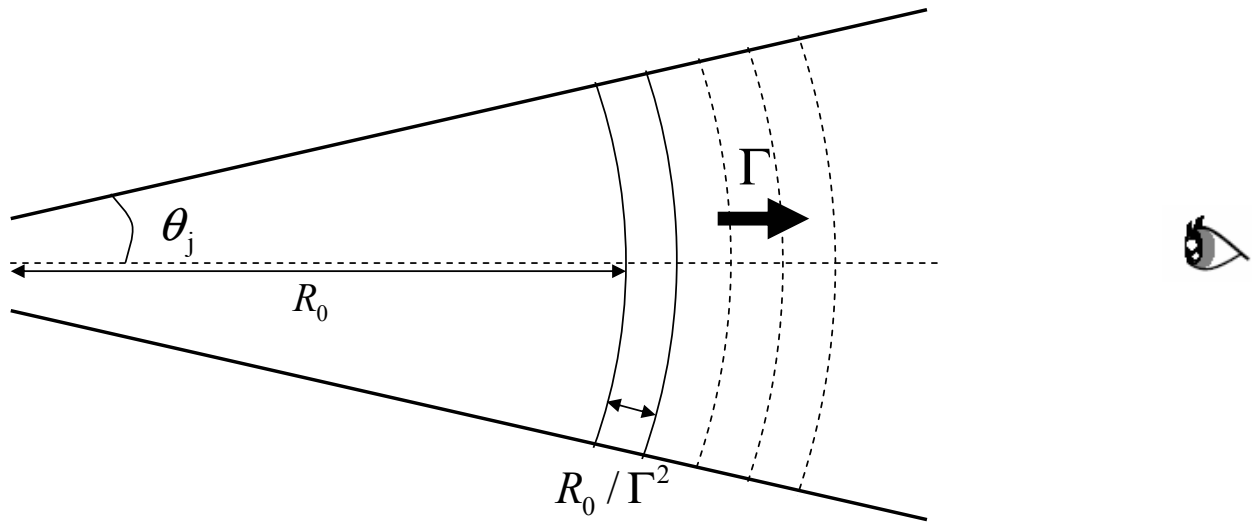}
\caption{Schematic picture of our model.
\label{fig:1}}
\end{figure}

Our model is summarized in Figure \ref{fig:1}.
The calculation starts at a radius $R=R_0$, where high-energy electrons
also start to be injected.
The quasi-steady outflow is modeled by identical shells
continuously ejected from $R=R_0$.
We consider a shell region of a constant width $W=R_0/\Gamma^2$
($R_0/\Gamma$ in the comoving frame)
that is moving outward with Lorentz factor
$\Gamma=1/\sqrt{1-\beta^2}$.
Our time-dependent numerical code \citep{asa11} can follow the evolution of
the electron energy distribution and photon production in the shell
with increasing radius $R$.
Our numerical code was developed for GRBs
so that the geometry of the jet is assumed to be a cone with a constant half-opening angle
$\theta_{\rm j}$, while the emission region of blazars has been frequently
modeled as a spherical blob or cylindrical flow
in previous studies.
Here, we assume a narrow cone with $\theta_{\rm j}=1/\Gamma$.
Given this opening angle, the transverse scale of the jet $R/\Gamma$ is
comparable to the radial scale in the comoving frame.
In this case, the curvature effect of the cone is not very important.
This geometry is not significantly different from spherical or cylindrical
emission zones.

We regard this thin shell as a homogeneous region
and particle distributions are assumed to be isotropic
in the comoving frame (one-zone approximation).
Our one-zone numerical code includes the effects of
electron cooling and injection to obtain the temporal
evolution of the plasma and photon production.
In this paper, we also add the acceleration and energy-diffusion
effects due to plasma-wave turbulence.

The evolution of the electron momentum distribution is described
by the Fokker--Planck equation as
%%%%%%%%%%%%%%%%%%%%%%%
\begin{eqnarray}
\frac{\partial f_{\rm e}(p,t)}{\partial t}
=\frac{1}{p^2} \frac{\partial}{\partial p}
\left[ p^2 D_{pp}(p) \frac{\partial f_{\rm e}(p,t)}{\partial p}
+ p^2 \langle \dot{p}
\rangle_{\rm cool} f_{\rm e}(p,t)
\right] - \frac{\dot{V}}{V} f_{\rm e}(p,t)
+\dot{f}_{\rm e,inj}(p,t),
\label{secF0}
\end{eqnarray}
%%%%%%%%%%%%%%%%%%%%%%%
where we have assumed an isotropic and homogeneous distribution
for the distribution function
$f_{\rm e}(p,t)$.
The electrons are assumed to be confined in a volume $V$.
The term with $\dot{V}/V$ expresses the density decrease due to
the volume expansion, where $\dot{V}$ is the volume expansion rate.
The effects of radiative/adiabatic cooling and particle injection are described
using the momentum loss rate $\langle \dot{p} \rangle_{\rm cool}>0$,
and $\dot{f}_{\rm e,inj}(p,t)$, respectively.
For ultra-relativistic particles, their energies
can be approximated as $\varepsilon_{\rm e}=c p$.
The homogeneous approximation allows us to describe
the total energy-distribution function as
$N_{\rm e}(\varepsilon_{\rm e},t)=4 \pi p^2 c^{-1} f_{\rm e}(p,t) V$.
Then, converting the diffusion coefficient $D_{pp}(p)$ into
$D(\varepsilon_{\rm e})=c^2 D_{pp}(p)$, eq. (\ref{secF0}) becomes
%%%%%%%%%%%%%%%%%%%%%%%
\begin{eqnarray}
\frac{\partial N_{\rm e}(\varepsilon_{\rm e},t)}{\partial t}
=\frac{\partial}{\partial \varepsilon_{\rm e}}
\left[ D(\varepsilon_{\rm e}) \frac{\partial N_{\rm e}(\varepsilon_{\rm e},t)}{\partial \varepsilon_{\rm e}}
\right]-
\frac{\partial}{\partial \varepsilon_{\rm e}}
\left[ \left( \frac{2 D(\varepsilon_{\rm e})}{\varepsilon_{\rm e}}-\langle \dot{\varepsilon_{\rm e}}
\rangle_{\rm cool} \right) N_{\rm e}(\varepsilon_{\rm e},t) \right]
+\dot{N}_{\rm e,inj}(\varepsilon_{\rm e},t),
\label{secF}
\end{eqnarray}
%%%%%%%%%%%%%%%%%%%%%%%
where $\langle \dot{\varepsilon_{\rm e}} \rangle_{\rm cool}$ is the energy loss rate,
and $\dot{N}_{\rm e,inj}(\varepsilon_{\rm e},t)$ is the total electron injection rate.
Electrons are gradually accelerated via scattering by turbulence.
If the average scattering frequency $\nu$
and the fractional energy change per scattering $\bar{\xi}$ are given,
the diffusion coefficient can be written as
%%%%%%%%%%%%%%%%%%%%%%%
\begin{eqnarray}
D(\varepsilon_{\rm e})=\frac{\bar{\xi}}{2} \varepsilon_{\rm e}^2 \nu.
\end{eqnarray}
%%%%%%%%%%%%%%%%%%%%%%%
A collision with a fluid element of velocity $\beta_{\rm d} \ll 1$
yields $\bar{\xi} \simeq 4 \beta_{\rm d}^2/3$.
The average velocity of turbulence may be determined by
the Alfv\'en velocity or the sound velocity.
A fluid with a relativistic temperature
(the sound speed is $c/\sqrt{3}$)
gives an extreme limit of $\bar{\xi} \simeq 2/3$.
Quasi-linear theory implies that
the collision frequency $\nu$ is proportional to
the gyration frequency $\Omega=e B c/\varepsilon_{\rm e}$ as
%%%%%%%%%%%%%%%%%%%%%%%
\begin{eqnarray}
\nu \equiv \frac{\pi}{4} \frac{k |\delta B^2 |_k}{B^2} \Omega,
\label{nu}
\end{eqnarray}
%%%%%%%%%%%%%%%%%%%%%%%
where $k \simeq eB/\varepsilon_{\rm e}$ is the wavenumber of turbulence
that resonates with the gyration frequency of the electrons \citep{bla87}.
The Fourier transform of the magnetic turbulence is assumed to be
a power-law function given by $|\delta B^2 |_k \propto k^{-q}$.
Then, as is well known,
the diffusion coefficient becomes a power-law function given by
%%%%%%%%%%%%%%%%%%%%%%%
\begin{eqnarray}
D(\varepsilon_{\rm e})=
\frac{\bar{\xi} \pi e c \varepsilon_{\rm e} k |\delta B^2 |_k}{8 B}
\equiv K \varepsilon_{\rm e}^q.
\label{Dep}
\end{eqnarray}
%%%%%%%%%%%%%%%%%%%%%%%
As shown in \citet{dun90},
the cross helicity state of the Alfv\'en waves
can affect the momentum diffusion coefficient.
However, here we simply extrapolate the above formula
for isotropic turbulences in the shell.

\begin{figure}[htb!]
\centering
\epsscale{1.0}
\plotone{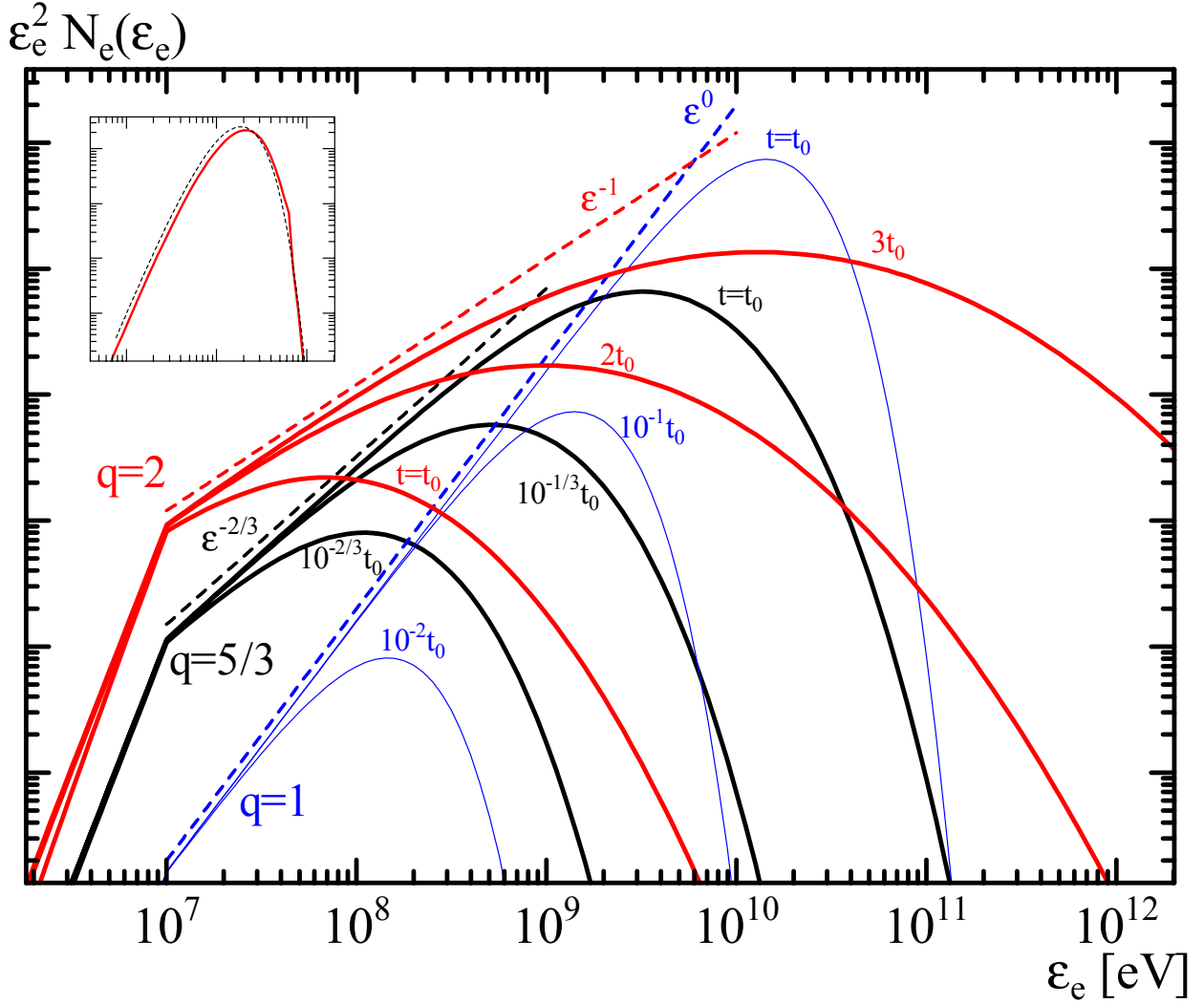}
\caption{Test calculations for the evolution of the electron energy distribution
with the Fermi-II acceleration for various indices $q$.
Here, we neglect the radiative cooling and volume expansion.
The labels for the dashed lines show a power-law form for
the energy distribution $N_{\rm e}(\varepsilon_{\rm e})$,
while the vertical axis (in arbitrary units)
is $\varepsilon_{\rm e}^2 N_{\rm e}(\varepsilon_{\rm e})$.
The inset figure shows a quasi-steady spectrum due to the
balance between the acceleration and synchrotron cooling
(arbitrary units). This agrees with the Maxwell-like function
(dashed line).
\label{fig:2}}
\end{figure}

In our numerical procedure,
for each time step, after the calculation for electron cooling,
the differential terms including $D(\varepsilon_{\rm e})$ in eq. (\ref{secF})
are evaluated with the MUSCL scheme with second-order accuracy
\citep{van79}
for first-order differentiation
and the central-difference method for second-order differentiation.
In Figure \ref{fig:2}, our test calculations neglecting the electron cooling
are shown.
Here, we continuously inject electrons at $10^7$ eV at a constant rate.
The acceleration timescale can be roughly written as
$t_{\rm acc} \sim \varepsilon_{\rm e}^2/2 D(\varepsilon_{\rm e}) \propto \varepsilon_{\rm e}^{2-q}$.
So, the steady-state solution provides
a power-law distribution $N_{\rm e}(\varepsilon_{\rm e}) \propto \varepsilon_{\rm e}^{1-q}$.
In Figure \ref{fig:2}, we normalize time by the acceleration timescale
for $10^{10}$ eV, $t_0 \equiv (10^{10} {\rm eV})^2/(2 D(10^{10} {\rm eV}))$.
The spectral evolution agrees with the acceleration timescale
and spectral index estimated above.
The cooling effect in this code is also checked.
\citet{sch85} provides time-dependent formulae of the electron distribution
under the simultaneous action of
synchrotron and IC radiation losses competing with Fermi-I and Fermi-II
accelerations.
Here, we simply show a steady-state case in Fermi-II models.
When the Fermi-II acceleration is balanced
by synchrotron energy losses, the electron distribution
becomes a Maxwell-like function $N_{\rm e} (\varepsilon_{\rm e}) \propto
\varepsilon_{\rm e}^2 \exp{
\left\{-(\varepsilon_{\rm e}/\varepsilon_{\rm c})^{3-q} \right\}}$
\citep{lef11}, which is identical to the steady-state solution of \citet{sch85}.
The inset figure in Figure \ref{fig:2}
shows a quasi-steady distribution after switching off
the injection but with acceleration for $q=2$.
In this test calculation, we consider only synchrotron cooling.
The distribution agrees with the analytical one (dashed line).

The energy source of the turbulence may be
dissipation of the bulk kinetic energy of the jet.
In those cases, jets are expected to be decelerated.
However, for simplicity, we assume a constant Lorentz factor $\Gamma$
throughout this paper.
If internal fluid motions exist in the jet at $R<R_0$,
the dissipation of the internal motions can be the energy source.
This model may validate the constant Lorentz factor in our simulations.
Regardless, we do not specify the dissipation source for the turbulence
and the second-order Fermi acceleration is phenomenologically treated
with the parameter $K$.

Hereafter, we fix the index $q$ in eq. (\ref{Dep})
at the Kolmogorov value $5/3$ for simplicity.
The diffusion coefficient should be determined by the characteristics of
the turbulences.
At the present time, we have no definite theory for the magnetic turbulence in blazar jets.
To reproduce the observed spectra, we will adjust diffusion coefficients below.
If the coefficient $K$ in eq. (\ref{Dep}) is larger than
the value $K_{\rm max} \sim (\gamma_{\rm max} m_{\rm e} c^2)^{1/3} \Omega$
estimated from eq. (\ref{nu}) with an extreme limit $\bar{\xi} k |\delta B^2|_k/B^2 \sim 1$
at $\varepsilon_{\rm e} \sim \gamma_{\rm max} m_{\rm e} c^2$,
it is physically unrealistic.
As will be seen below, the values of $K$ we adopt are safely smaller than
$K_{\rm max} \sim (\gamma_{\rm max} m_{\rm e} c^2)^{-2/3} e B c
\simeq 1.4 \times 10^4 (B/0.1~\mbox{G}) (\gamma_{\rm max}/10^6)^{-2/3}
~\mbox{eV}^{1/3}~\mbox{s}^{-1}$.

For simplicity, the electron injection is assumed to be
monoenergetic;
the electron Lorentz factor at injection will be fixed to $\gamma'_{\rm inj}=100$.
Hereafter,
the quantities in the comoving frame are denoted with primed characters.
As the shell outflows, the injected electrons are gradually accelerated
following eq. (\ref{secF}).
As the emission region flows outward, the volume increases as $V' \propto R^2$
in this conical geometry.
Adiabatic cooling is taken into account with the same method
as \citet{asa11}, in which the electron energy decreases
as $\varepsilon'_{\rm e} \propto V'^{-1/3}$ in the ultra-relativistic limit.
The electrons remaining in the shell cool adiabatically, and
the emission will cease as the shell expands even if
electrons do not escape from the shell.

We do not include the effect of the electron escape
in this paper as shown in eq. (\ref{secF0}).
This is a critical process for obtaining the electron spectrum,
as is well known.
In our quasi-steady outflow model depicted in Fig. \ref{fig:1},
the electron escape is equivalent to the electron transfer
between shells. Our one-zone approximation is not optimized
for the electron transfer. However, the isotropic diffusion we assumed
may allow us to neglect the escape, because the escape rate
from a shell may be almost equal to the incoming rate from the adjoining shells.
Given the mean free path $l_{\rm m}=c/\nu$, the spatial diffusion coefficient
can be approximated as $D_{xx}=l_{\rm m} c/3$.
Then, the diffusion length in the dynamical timescale $W'/c$
becomes $\left< \delta x' \right> =\sqrt{D_{xx} W'/c}$, which implies
%%%%%%%%%%%%%%%%%%%%%%%
\begin{eqnarray}
\frac{\left< \delta x' \right>}{W'}=\sqrt{\frac{\bar{\xi} \varepsilon^{\prime 1/3}_{\rm e}
c}{6 W' K'_0}}
\simeq 0.7 \bar{\xi}^{1/2} \left( \frac{W'}{10^{16}~\mbox{cm}} \right)^{-1/2}
\left( \frac{K'_0}{10^{-2}~\mbox{eV}^{1/3}~\mbox{s}^{-1}} \right)^{-1/2}
\left( \frac{\varepsilon'_{\rm e}}{10^{12}~\mbox{eV}} \right)^{1/6},
\end{eqnarray}
%%%%%%%%%%%%%%%%%%%%%%%
for $q=5/3$.
For a conservative value of $\bar{\xi} \ll 1$,
the spatial diffusion is not sufficient.
This also supports neglecting the escape effect in our model.

The average magnetic field should decay with radius, unless some kind of
amplification mechanism is at work.
In this paper, we assume a power-law evolution as
$B'=B_0 (R/R_0)^{-1}$, which implies conservation of magnetic energy.

In each time step, photons are produced in the shell
with a rate that is consistent with the electron cooling rate.
The photon production processes we adopt are
synchrotron and IC emission.
The Klein--Nishina effect on IC emission is fully included
in our numerical method.
The photon density is evaluated with the homogeneous approximation
taking into account the photon escape from the shell
\citep[see][for details]{asa11}.
We adopt this spectral density of photons
to estimate the seed photons for IC scattering.
Our one-zone approximation does not solve the radiative transfer
in the steady outflow.
Therefore, the photons that escape are not counted as the seed photons
for IC scattering.
Such photons may contribute to the seed photons in regions outside
the original shell.
However, we take into account only the photons remaining in the shell.
This problem in our method may be absorbed by the uncertainty
in the model setting (simplified geometry, electron injection etc.).
The photon absorption via $\gamma \gamma$ collision,
secondary electron--positron pair injection, and synchrotron
self-absorption are also included in our code.
However, those effects are not so important in our examples below.

The evolution of accelerated particles and photon production
in a shell are computed with the time-dependent method, as we explained above.
Considering the curvature effect, Doppler boosting due to the relativistic
bulk motion of the shell, and the opening angle $\theta_{\rm j}$,
we can estimate the arrival time and energy of photons
escaping from the shell for an observer.
Based on those outputs, the temporal evolution of the photon spectrum
emitted from ``one shell'' can be obtained.
High-energy photons can be absorbed via $\gamma \gamma$ collisions
with the EBL during propagation in the intergalactic medium.
To obtain the spectra seen by observers, we adopt the model in \citet{kne04}
for the EBL evolution.

The central engine may continuously eject shells that
emit photons.
Photons escaping from different shells can arrive at an observer
simultaneously.
We can model the temporal evolution of blazar emission by adding
the contributions of such shells with different launch times.
If we change the model parameters for each shell,
various models including steady emission will be realized.
To model steady emission from a steady flow,
we assume that identical shells are continuously ejected at $R=R_0$
with a time step of $R_0/(c \beta \Gamma^2)$.
The steady spectrum for an observer is comprised of the contributions
of all shells at $R \geq R_0$.
The time-integrated spectrum from one shell provides the average spectral
shape from the steady flow.
The steady spectral flux is easily obtained by
dividing the time-integrated spectrum
emitted from one shell by the shell ejection time step $R_0/(c \beta \Gamma^2)$.

For the steady emission model, there are six model parameters:
the initial radius $R_0$, the radius where the injection and acceleration cease
$R_{\rm c}$, the bulk Lorentz factor $\Gamma$, the initial magnetic field $B_0$,
the injection rate $\dot{N'_{\rm e}}$, and the diffusion coefficient $K'$.
These are the minimum parameters required in our model.
In \S \ref{sec:Mrk}, we will additionally consider the radial evolution
of $\dot{N'_{\rm e}}$ and $K'$.
In this case, the power-law indices are introduced as two additional parameters.
The number of parameters is not many compared with previous models.
For example, the model parameters for Mrk 421 in \citet{mrk421wide},
a one-zone synchrotron self-Compton (SSC) model, is 11.

\section{1ES 1101-232}
\label{sec:1ES}

First, we adopt the Fermi-II acceleration model
for the TeV blazar 1ES 1101-232 \citep{1101}.
The detection of TeV gamma-rays from this high redshift ($z=0.186$) object
piqued interest in light of the constraint on the EBL \citep{1101EBL}.
The {\it Fermi} telescope provided
a stringent upper limit in the GeV energy range \citep{ner10}.
This implies a very hard spectrum from GeV to TeV.

\begin{figure}[htb!]
\centering
\epsscale{1.0}
\plotone{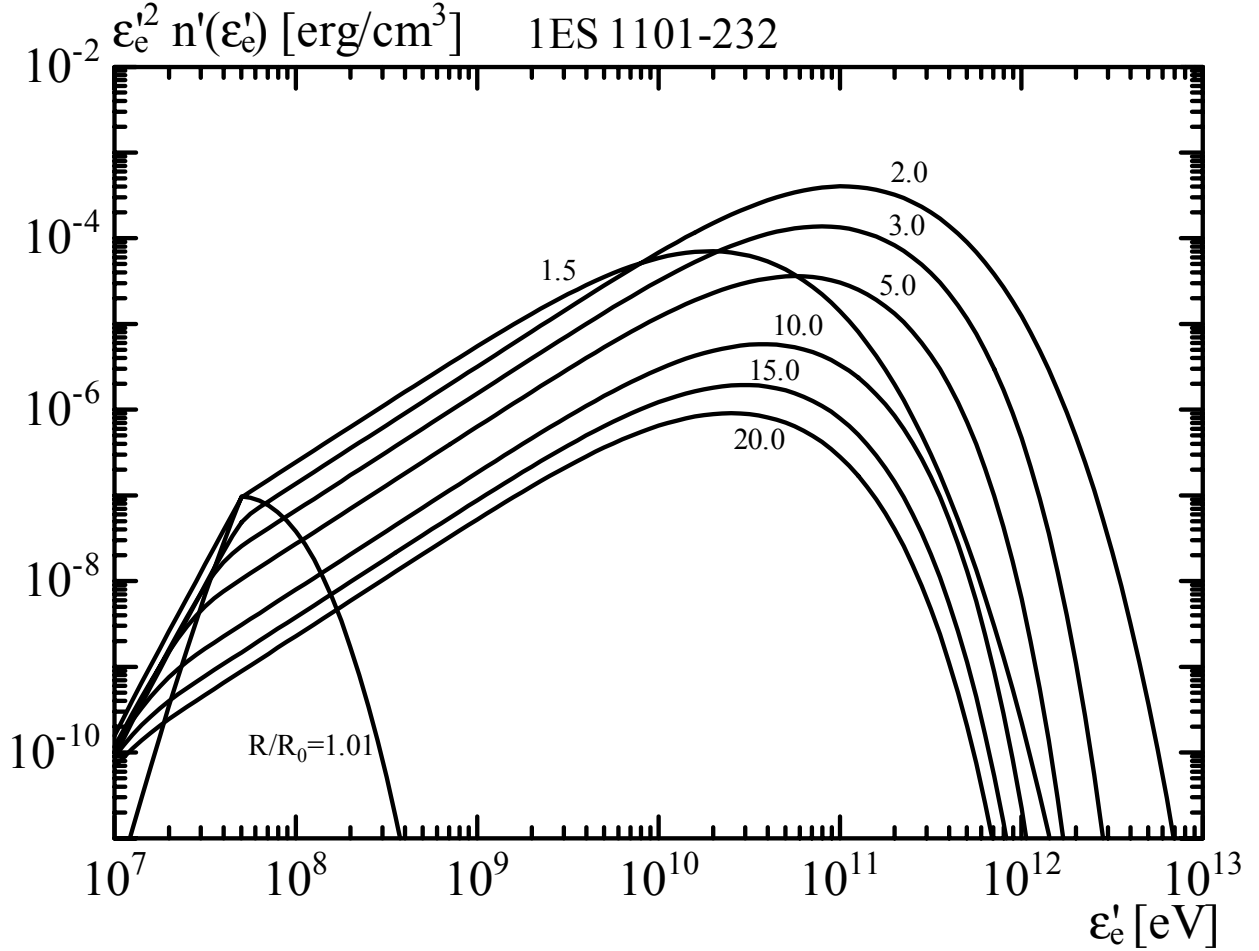}
\caption{Evolution of the electron energy distribution for 1ES 1101-232
\label{fig:3}}
\end{figure}

The model parameters for this blazar are $\Gamma=25$, $B_0=0.03$ G,
and $W'=R_0/\Gamma=2.8 \times 10^{16}$ cm.
We inject electrons with a constant rate
$\dot{N}'_{\rm e}=\dot{N}_0=
1.5 \times 10^{46}~\mbox{s}^{-1}$ in spherically symmetric evaluation
over a timescale of $\Delta T'_{\rm inj}=W'/c$ in the shell frame.
This implies that the electron injection ceases at $R=2 R_0$.
In this injection timescale, turbulence in the shell accelerates electrons
with the diffusion coefficient
$K'=4.3 \times 10^{-3}~\mbox{eV}^{1/3}~\mbox{s}^{-1}$.
After the end of the electron injection, we assume that the
turbulence is terminated as well, so electrons cool
monotonically via radiation and adiabatic expansion.

Figure \ref{fig:3} shows the evolution of the electron energy distribution
in the shell frame.
As the electron injection and acceleration proceed,
the electron energy density grows and achieves a maximum
at $R=2R_0$.
The $\varepsilon_{\rm e}'^2 n'(\varepsilon_{\rm e}')$-spectrum has a maximum
at $\sim 10^{11}$ eV,
where $n'(\varepsilon_{\rm e}') \equiv N'_{\rm e}(\varepsilon_{\rm e}')/V'$.
This peak energy is determined by the duration time
of the acceleration, which corresponds to the acceleration timescale
of this energy.
After the end of the electron injection and acceleration,
the shell expansion causes the density drop,
and adiabatic cooling lowers the electron energy.
Thus, the spectral peak energy shifts to lower energies as the shell expands.
The effect of the radiative cooling is seen as
the growth of the sharpness of the spectral cut-off
above the peak energy.

\begin{figure}[htb!]
\centering
\epsscale{1.0}
\plotone{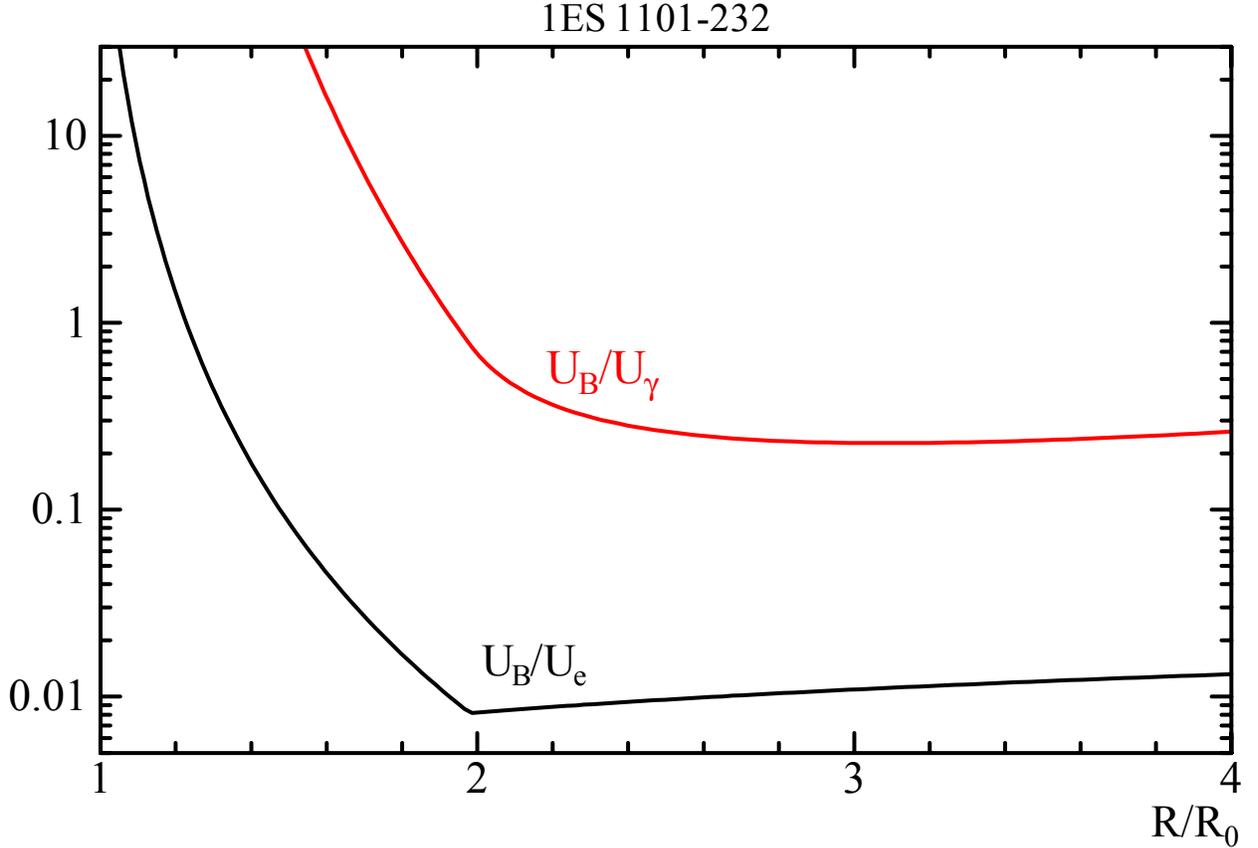}
\caption{Evolution of the energy density ratios in a shell
in the model for 1ES 1101-232.
The energy densities of photons, electrons,
and magnetic fields in the shell frame are denoted
by $U_\gamma$, $U_{\rm e}$, and $U_B$, respectively.
\label{fig:4}}
\end{figure}

In this parameter set,
the radiative cooling is not so efficient
that the photon energy density is always lower than
the electron energy density,
as indirectly shown in Figure \ref{fig:4}.
Nonetheless, the photon energy density
overtakes the magnetic energy density in the later phase,
which leads to sufficient SSC emission.

\begin{figure}[htb!]
\centering
\epsscale{1.0}
\plotone{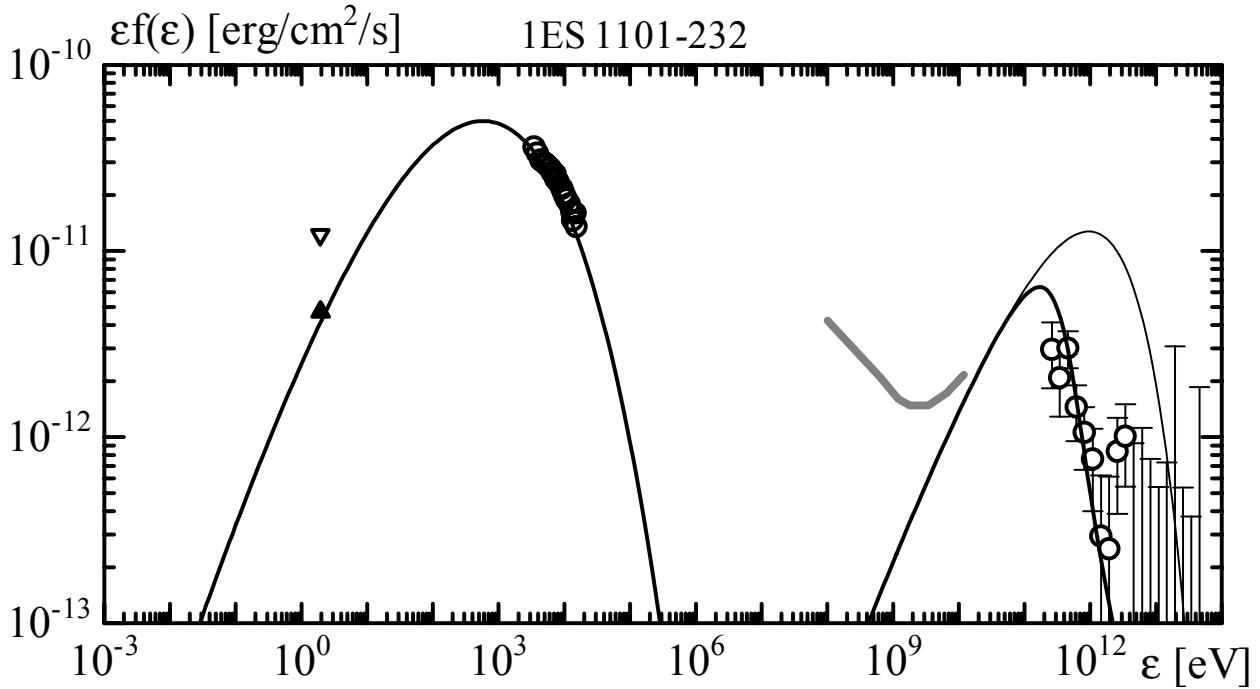}
\caption{Steady photon spectrum for 1ES 1101-232.
The data points are from \citet{1101} and
the {\it Fermi} upper limit in \citet{ner10}
is also shown with the bold gray line.
The thin line for the IC component is the spectrum without
the absorption effect due to the EBL.
\label{fig:5}}
\end{figure}

The steady-state spectrum obtained from our model
is shown in Figure \ref{fig:5}.
Our simple assumption (constant injection and diffusion coefficient)
succeeds in reproducing the observed hard spectrum and
avoiding the {\it Fermi} upper limit.
The hard electron spectrum due to the Fermi-II acceleration
naturally leads to this hard spectrum.

\section{Steady Emission in Mrk 421}
\label{sec:Mrk}

The Fermi-II acceleration model can naturally explain
the hard-spectrum blazar as shown in the previous section.
If this acceleration mechanism is universal in the quasi-steady emission
from blazars, relatively softer spectra for other ordinary blazars
should be also fitted by this model.
However, the hard spectra obtained from the simplest model
apparently contradict the observed one.
In order to overcome this problem, we consider the temporal (equivalently radial)
evolution of the electron injection rate and diffusion coefficient.

As a representative example of blazars, we consider Mrk 421 at $z=0.031$,
whose broadband spectrum from the radio to TeV is one of the most
precisely observed spectra.
Here, we adopt the spectrum obtained from the 4.5 month long multi-frequency campaign
\citep[2009 January 19 to 2009 June 1;][]{mrk421wide}. During this campaign,
Mrk 421 showed low activity and
relatively small flux variations at all frequencies.
Thus, this data set can be used to study steady emission from blazars.
In \citet{mrk421wide}, to fit the obtained spectrum by leptonic models,
an electron distribution of three power-law functions (namely two breaks)
is required.
This may be because the spectral shape around the peak energy
from optical to X-ray bands
is too broad for single-break models.
While the origin of such spectral breaks is unknown,
the time-dependent model may provide us a new possible picture
for this blazar.

For electrons injected at a later phase,
the effective duration of the acceleration becomes
shorter than that of the electrons injected initially.
Such electrons injected later remain in
the low-energy regime.
Therefore, an increase in the injection rate for
a finite injection timescale leads to a softer electron
spectrum than that with a constant injection rate.
The diffusion coefficient may also evolve with time.
A decrease of the diffusion coefficient makes
electrons injected later remain in the low-energy regime.
However, a too rapid decline of $K'$ results in
a too low maximum energy of the electrons.
In order to reproduce the observed spectrum, hereafter,
we adjust the evolution of $\dot{N}'_{\rm e}$,
while the evolution of $K'$ is fixed as $K' \propto R^{-1}$
for simplicity.

\subsection{Simple SSC model}
\label{421SSC}

\begin{figure}[htb!]
\centering
\epsscale{1.0}
\plotone{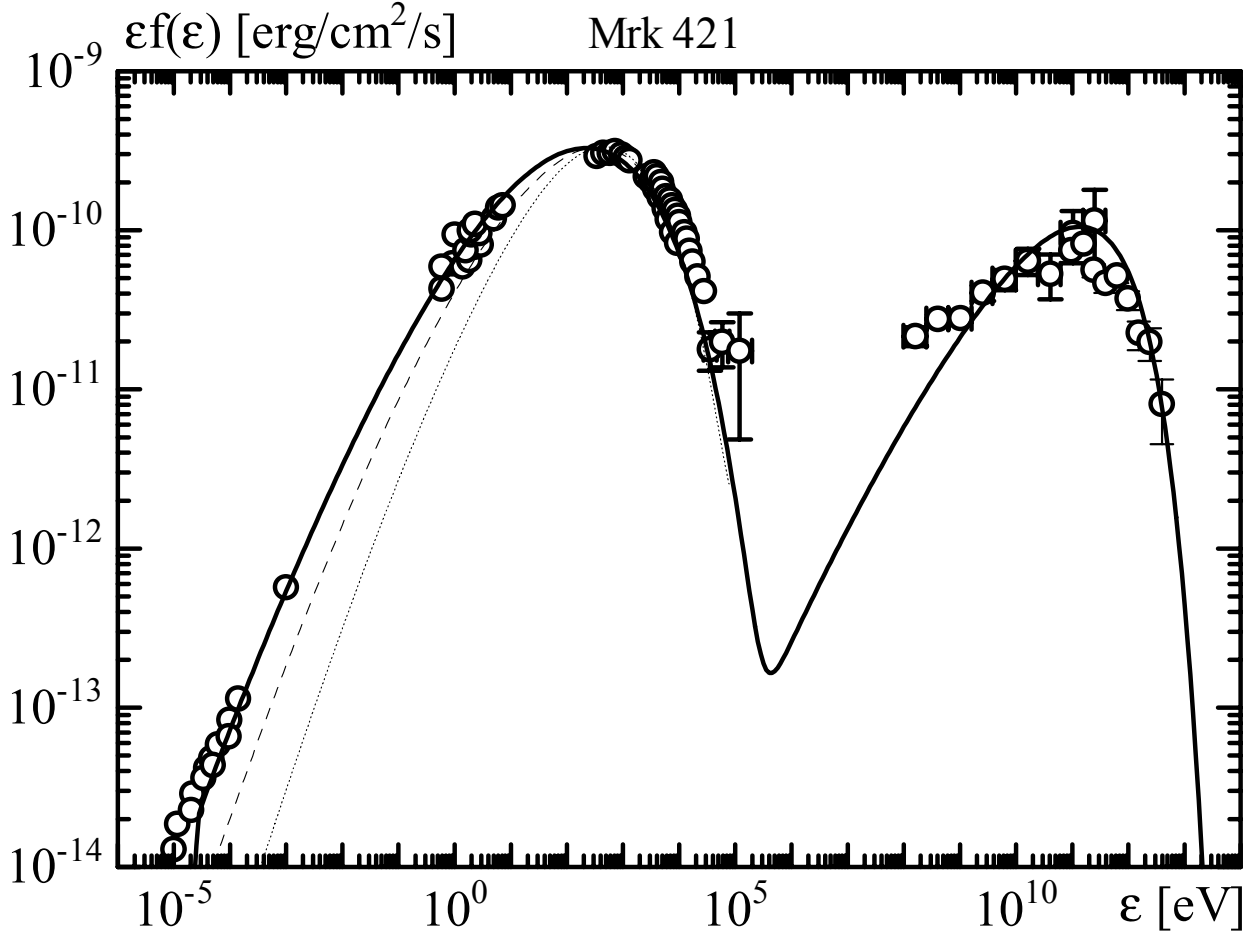}
\caption{Steady photon spectrum for the simple SSC model of Mrk 421
(see \S \ref{421SSC}).
The data points are partially extracted from the data of the 4.5 month campaign
\citep{mrk421wide}.
For reference, synchrotron spectra with
different parameter evolutions,
$\dot{N}'_{\rm e} \propto R^7$ and $K' \propto R^0$ (thin dashed line)
and $\dot{N}'_{\rm e} \propto R^0$ and $K' \propto R^{-1}$ (thin dotted line)
are plotted.
\label{fig:6}}
\end{figure}

Figure \ref{fig:6} shows the result obtained from our model
with the temporal evolution of the injection rate and
diffusion coefficient.
The model parameters are $\Gamma=15$, $B_0=0.13$ G,
and $W'=R_0/\Gamma=1.0 \times 10^{16}$ cm.
The duration time of the electron injection and acceleration
is assumed to be $\Delta T'_{\rm inj}=2 W'/c$ (end at $R=3 R_0$),
which is longer than the assumption in 1ES 1101-232
to enhance the effects of the temporal evolution.
In this duration time, the injection rate is assumed to evolve as
$\dot{N}'_{\rm e}=\dot{N}_0 (R/R_0)^7$, where
$\dot{N}_0=9.8 \times 10^{43}~\mbox{s}^{-1}$.
Similarly, the diffusion coefficient evolves as
$K'=K_0 (R/R_0)^{-1}$, where
$K_0=1.3 \times 10^{-2}~\mbox{eV}^{1/3}~\mbox{s}^{-1}$.

The synchrotron component is well reproduced by this model.
An advantage of this model is that the curved spectral feature
is naturally explained by the power-law evolution of the injection and diffusion,
while the usual shock acceleration models need breaks at ad hoc energies
in the injection spectrum.

\begin{figure}[htb!]
\centering
\epsscale{1.0}
\plotone{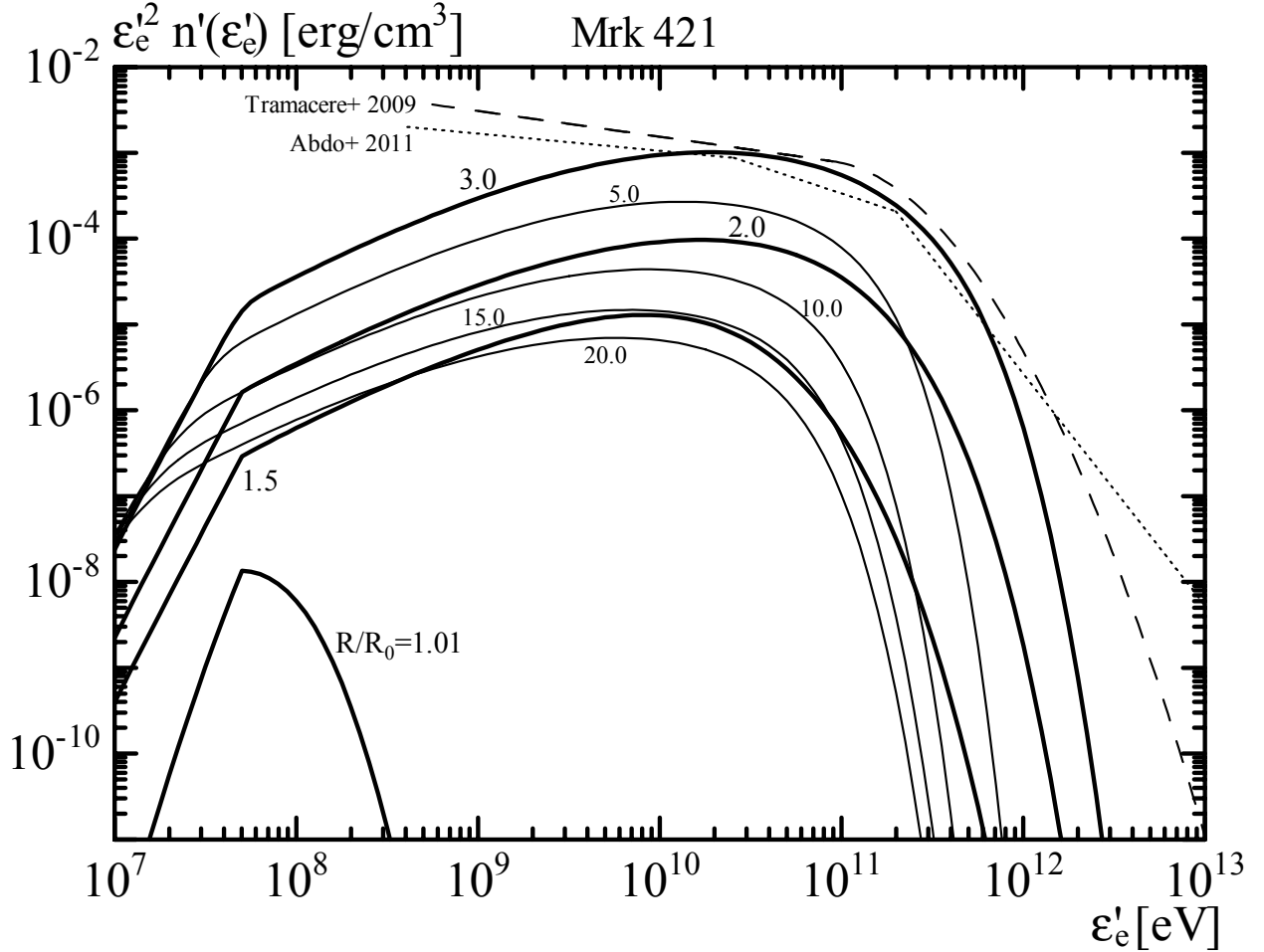}
\caption{Evolution of the electron energy distribution for
the model of Mrk 421 in Fig. \ref{fig:6}.
The electron spectra during the acceleration process 
are denoted by thick lines, while those after the end of the acceleration
are denoted by thin lines.
The spectral shapes of other one-zone leptonic models
are plotted (in arbitrary units) for reference.
The thin dotted line is the DBP model
in \citet{mrk421wide} and the thin dashed line is the log-parabolic
model in \citet{mrk421swift}.
\label{fig:7}}
\end{figure}

The curved photon spectrum is a direct consequence of the
curved electron spectrum, as shown in
Figure \ref{fig:7}.
The electron spectra are softer than the case in 1ES 1101-232
owing to the temporal evolution of the electron injection.
Just above $\varepsilon'_{\rm e}=\gamma'_{\rm inj} m_{\rm e} c^2$,
the electron spectral index is about 1.06,
but the spectrum gradually becomes softer with increasing energy.
After the acceleration ceases, the electron spectra show
a sharper cut-off due to the radiative cooling (see the thin lines in Figure \ref{fig:7}).
For reference, we also plot the analytic model spectra
in \citet{mrk421wide} and \citet{mrk421swift}.
The double broken power-law (DBP) model in \citet{mrk421wide}
has breaks at $2.6 \times 10^{10}$ eV and $2.0 \times 10^{11}$ eV
with indices of $2.2$, $2.7$, and $4.7$ from low to high energy.
The model in \citet{mrk421swift} is a combination
of a power-law at low energies (index $2.3$) and a log-parabolic
high-energy branch:
%%%%%%%%%%%%%%%%%%%%%%%
\begin{eqnarray}
n'(\varepsilon'_{\rm e}) \propto
\varepsilon'^{-2.3-0.75 \ln (\varepsilon'_{\rm e}/\varepsilon_{\rm j})}_{\rm e},
\end{eqnarray}
%%%%%%%%%%%%%%%%%%%%%%%
where the intersection of the two functions
is at $\varepsilon_{\rm j}=8.9 \times 10^{10}~\mbox{eV}$.
Note that this log-parabolic model was adopted to fit the
spectral data of 22-04-2006, while the DBP model
is for the same data set as ours. Those two analytic models have
similar shapes
to ours between $10^{10}$ eV and $10^{12}$ eV so that all the
models can fit the synchrotron component around the peak.

In the two models with analytic functions,
the soft spectra below $\sim 10^{10}$ eV
are advantageous to fit the GeV spectrum,
while our model flux is significantly lower
than the flux data obtained with {\it Fermi}.
It may be hard to make such a soft spectrum in this energy range
in Fermi-II models.
However, the power-laws for the two analytic models
below $10^{9}$ eV are too soft to reconcile with
the radio spectrum. Hence, the analytic models require another break
or low-energy cut-off below $10^{9}$ eV.
To discriminate these distributions, 
future infrared and submillimeter observations will be interesting.

The hard electron spectrum in our model
yields the radio emission that agrees with the observed spectrum.
In most of preceding models based on shock acceleration,
abundant low-energy electrons for a power-law index of 1.5-2.0
make the synchrotron self-absorption frequency fall in the submillimeter range.
Hence,
another component, such as synchrotron self-absorbed emission
from inhomogeneous jets \citep[e.g.][]{kon81} etc.,
has been required to reconcile the radio observations.
In contrast, the electron spectrum in the low-energy part
is rather hard with a power-law index of approximately $1.06$
even though we increased the low-energy particles with an evolution of the injection.
So, synchrotron self-absorption is negligible in our model.

We now confirm this statement analytically.
At $R=3 R_0$, the spectral density is
$n'(\varepsilon'_{\rm e}) \simeq 2200~\mbox{erg}^{-1}~\mbox{cm}^{-3}$
at $\varepsilon'_{\rm e}=\gamma'_{\rm inj} m_{\rm e} c^2$.
If we denote this as $n'(\varepsilon'_{\rm e})=C \varepsilon'^{-1}_{\rm e}$,
$C \simeq 0.18~\mbox{cm}^{-3}$, which is comparable to
the density $\sim 1~\mbox{cm}^{-3}$ obtained with the time integrated
number of electrons, $3^8 R_0 \dot{N}_0/8 c \Gamma$
(note $dt'/dR \simeq 1/c \Gamma$) and volume, $V'=4 \pi (3 R_0)^2 W'$.
The formula in \citet{ryb79} gives the optical depth due to synchrotron
self-absorption as
%%%%%%%%%%%%%%%%%%%%%%%
\begin{eqnarray}
\tau_{\rm SSA}=1.3 \times 10^{-2} \left( \frac{C}{0.18~\mbox{cm}^{-3}}
\right) \left( \frac{B'}{0.1~\mbox{G}} \right)^{3/2}
\left( \frac{\varepsilon'}{10^{-5}~\mbox{eV}} \right)^{-5/2}
\left( \frac{W'}{10^{16}~\mbox{cm}} \right),
\end{eqnarray}
%%%%%%%%%%%%%%%%%%%%%%%
or the break photon energy, defined as $\tau_{\rm SSA}(\varepsilon'_{\rm a})=1$,
becomes
%%%%%%%%%%%%%%%%%%%%%%%
\begin{eqnarray}
\varepsilon'_{\rm a}=1.8 \times 10^{-6} \left( \frac{C}{0.18~\mbox{cm}^{-3}}
\right)^{2/5} \left( \frac{B'}{0.1~\mbox{G}} \right)^{3/5}
\left( \frac{W'}{10^{16}~\mbox{cm}} \right)^{2/5}~\mbox{eV}.
\label{absfreq}
\end{eqnarray}
%%%%%%%%%%%%%%%%%%%%%%%
As shown in Figure \ref{fig:7}, the low-energy electron density in our
model is much less than the extrapolations of the analytic models.
As we have discussed,
this is one of the reasons why the self-absorption frequency is relatively low.
Distinct from the assumption in the above analytical estimate
of eq. (\ref{absfreq}),
the electron distribution has a break at $\gamma_{\rm e}=\gamma_{\rm e,inj}=100$,
so that the above break energy would decrease.
Thus, the spectral break at $\varepsilon \sim 10^{-5}$ eV
is mainly due to the break in the electron spectrum
rather than the absorption effect.
The actual $\gamma_{\rm e,inj}$ may be smaller than what we assumed
($\gamma_{\rm e,inj}$ has been set as 100 to save computational costs).
So, the electron spectrum in our model
can comprehensively explain the spectrum from the radio
to the X-ray without introducing a minimum Lorentz factor
$\gamma_{\rm e,min}$.

In our model, while the diffusion coefficient decreases,
a sharp rise in the injection rate ($\dot{N}'_{\rm e} \propto R^7$)
is required, which may seem unnatural.
The electron injection rate is determined by short-wavelength turbulence
that resonates with the gyro motion of low-energy electrons.
Such waves may have a different evolution from the turbulence
that accelerates high-energy electrons.
While the long waves are produced by large-scale instabilities,
such as the Kelvin--Helmholtz instability etc.,
the origin of the short waves may be the cascade of the long waves.
In this case, the efficiency of the electron injection may grow later
relative to the development of the long waves.
Another possibility is that
the low-energy threshold of the electrons in the acceleration process
decreases gradually.
Let us consider electrons at energies of the cut-off tail in the Maxwellian distribution.
When the minimum wavelength is relatively long,
only higher energy electrons can be injected into the acceleration process.
If the minimum wavelength gradually decreases as the cascade proceeds,
lower energy electrons are also injected.
This mechanism may cause a sharp rise in the injection rate
retracing the cut-off shape in the Maxwellian distribution.

We have assumed the evolutions of $\dot{N}'_{\rm e} \propto R^7$
and $K' \propto R^{-1}$. If either $\dot{N}'_{\rm e}$ or $K'$ is constant,
as shown in Figure \ref{fig:6},
the synchrotron spectrum becomes narrower than the observations.
The broad peak represented by the X-ray and IR-optical data points
is achieved by the combination of this evolution.
Of course, our example of the parameter evolutions may not be a unique solution.
On the other hand, we find that the X-ray spectrum shape can be solely fitted
without this evolution, if we neglect the IR-optical and radio data.
The X-ray spectral shape is determined by the high-energy cut-off shape
of the electron spectrum, which may be controlled by the diffusion
process in momentum space and radiative cooling
rather than the parameter evolution.

Our parameter choice reproduces the flux level of the IC component as well.
However, the observed flux at $\sim 100$ MeV is
significantly higher than the model spectrum.
The spectrum obtained with {\it Fermi} is relatively flat compared with
the synchrotron spectrum.
The steady SSC spectrum obtained with our time-dependent model
is hard to reconcile with the {\it Fermi} data.

\subsection{SSC+EIC model}
\label{421EIC}

The simplest method to fit to the GeV flux is an introduction of
another emission region that contributes to this energy range.
Such two-zone models have been discussed by several authors
such as \citet{ghi05}.

Here, we consider another possibility,
the effect of an external photon field, to reproduce
the $100$ MeV--GeV flux in Mrk 421.
While external photons are indispensable to explain IC components
of flat spectrum radio quasars (FSRQs),
BL Lac objects have been fitted without external photons.
For FSRQs,
optical photons from broad-line regions are a candidate
for the external photon field.
However, the average electron energy in BL Lac objects
is much higher than that in FSRQs so that the Klein--Nishina effect
makes the contribution of the external optical photons negligible.
Moreover, the typical energy range of the IC-scattered optical photons
becomes much higher than the GeV energy range.
Here, we consider external radio photons, which may come
from compact radio lobes as seen in young radio-loud AGNs
\citep{sne04}.

We consider an external photon field whose spectral peak
in the $\varepsilon f(\varepsilon)$-diagram is $10^{-6}$ eV (240 MHz).
This corresponds to $\sim 10^{-6} \Gamma$ eV
$\sim 10^{-5}$ eV in the shell frame,
which is safely high enough to avoid synchrotron self-absorption
[see eq. (\ref{absfreq})].
Since we have no definite model for the spectrum,
the Band function \citep{ban93}, smoothly joined power laws,
is adopted here.
The low- and high-energy photon indices
(defined as $-d \ln f(\varepsilon)/d \ln \varepsilon+1$) are chosen to be
$-1$ and $2.5$, respectively.
The total luminosity is $L_{\rm ex}=4.9 \times 10^{38}~\mbox{erg}~\mbox{s}^{-1}$.
%If the source can be approximated as a point source,
%the photon energy density is $U_{\rm ex}=L_{\rm ex}/\pi R^2 c$ at a distance $R$.
When photons are isotropically distributed,
the photon energy density in the comoving frame of the jet
is $4 U_{\rm ex} \Gamma^2/3$ \citep{der02}.
However, the isotropic approximation %and the point source are conflictive.
may not be accurate.
So we neglect the numerical coefficient,
and assume the comoving energy density to be
%%%%%%%%%%%%%%%%%%%%%%%
\begin{eqnarray}
U'_{\rm ex}=\Gamma^2 \frac{L_{\rm ex}}{\pi R^2 c}.
\end{eqnarray}
%%%%%%%%%%%%%%%%%%%%%%%
The spectral shape is simply shifted by a factor of $\Gamma$ in the shell frame.
Based on this photon distribution in the shell frame,
we calculate the contribution of external IC (EIC).
Of course, our time-dependent code can wholly take into account
the non-linearities of the cooling processes \citep{zac12}.
For simplicity, we assume isotropic emission in the shell frame,
although the external photons may be beamed in this frame.
Therefore, the contribution of the external photons
is simply taken into account by adding the boosted external photons
to the photon field in the shell frame.
The external component is intrinsically
indistinguishable from the internal synchrotron/IC photons.

\begin{figure}[htb!]
\centering
\epsscale{1.0}
\plotone{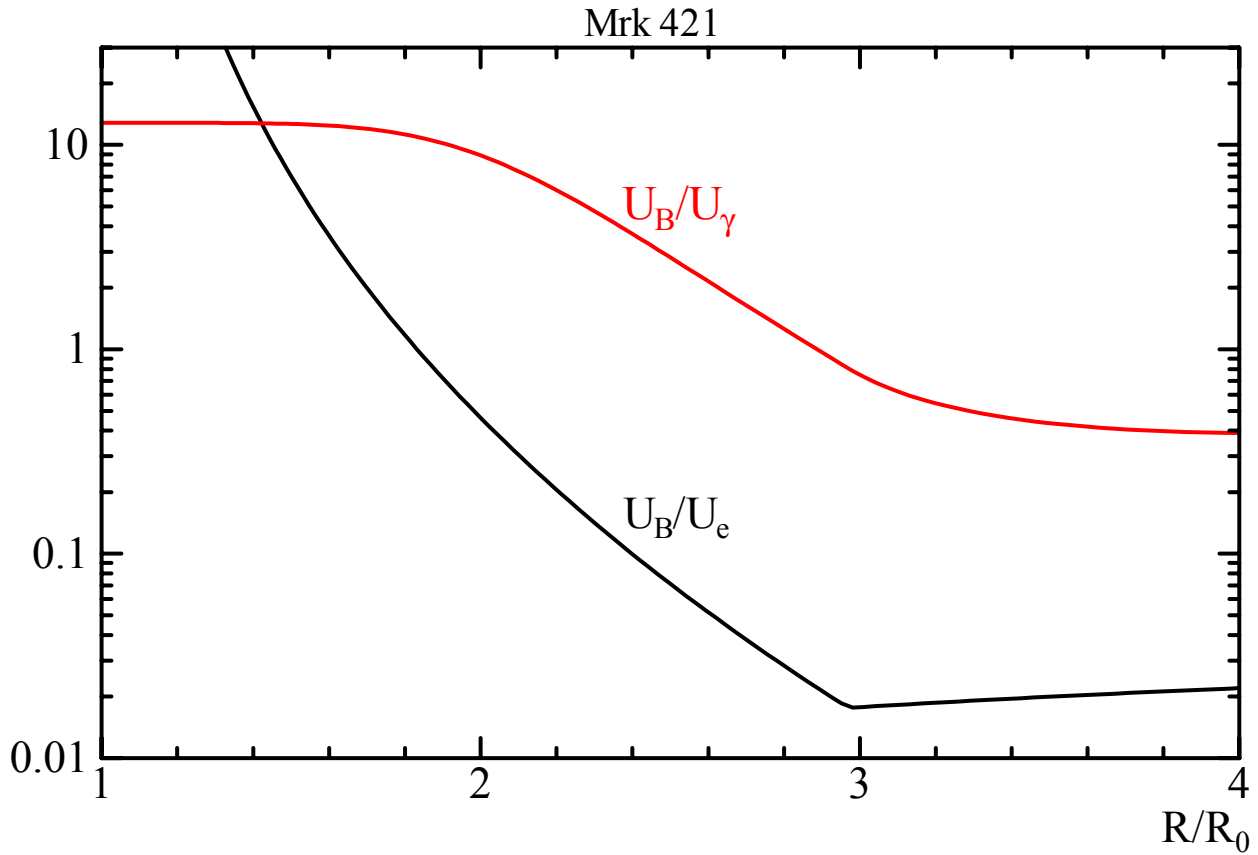}
\caption{Evolution of the energy density ratios in a shell in
the model for Mrk 421 with external photons
(see \S \ref{421EIC}).
The label notations are the same as in Fig. \ref{fig:4}.
\label{fig:8}}
\end{figure}

As shown in Figure \ref{fig:8}, the photon energy density
is initially dominated by the external photons
so that the ratio $U_B/U_\gamma$ is almost constant.
As the electron injection proceeds,
photons produced in the shell becomes predominant,
as seen at $R > 2 R_0$,
and its energy density overtakes the magnetic one.
Similarly to 1ES 1101-232, the emission efficiency is so low
that most of the electron energy is not released as radiation.

\begin{figure}[htb!]
\centering
\epsscale{1.0}
\plotone{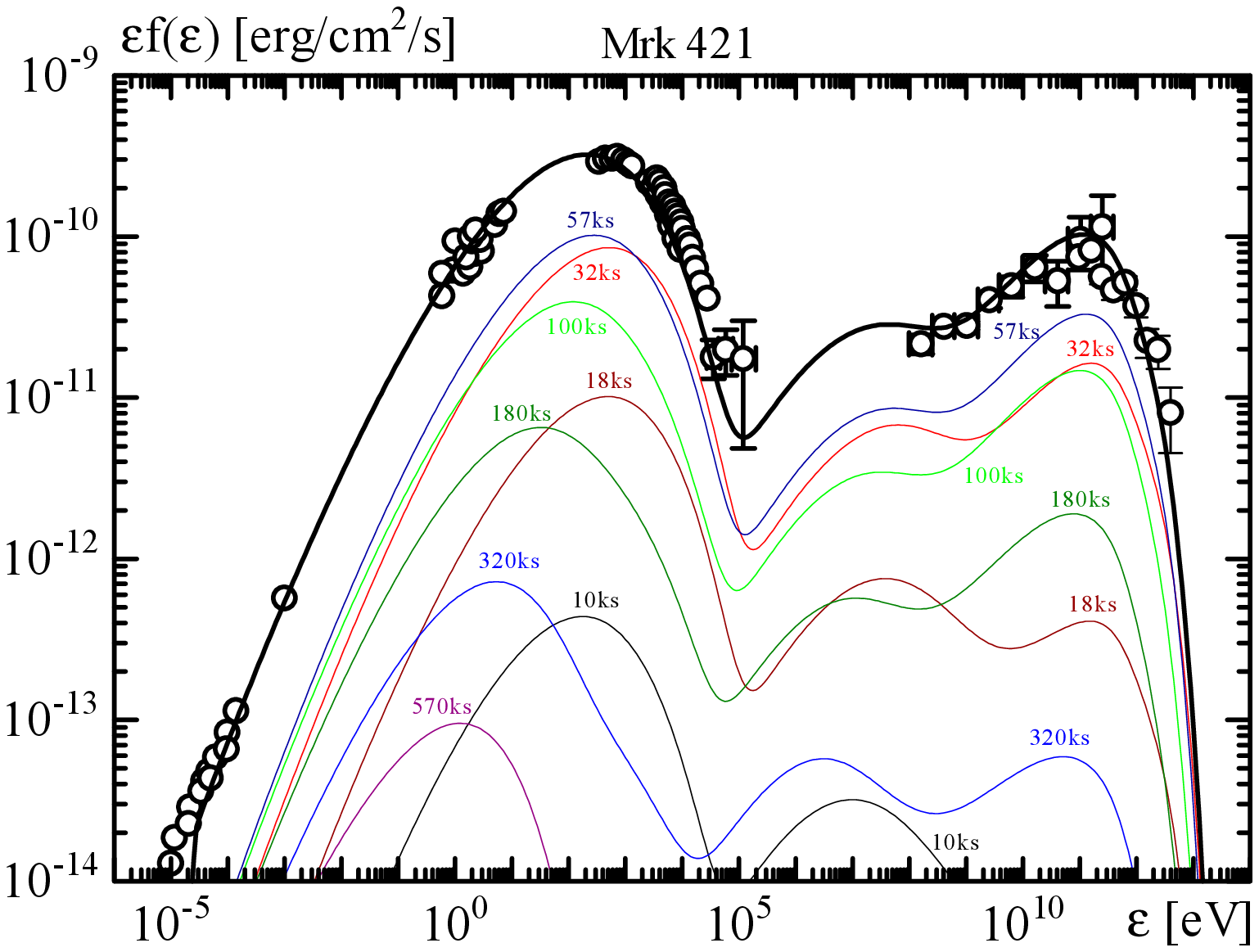}
\caption{Steady photon spectrum (thick) for the model of Mrk 421 with external photons
(see \S \ref{421EIC}).
Thin lines show the evolution of the photon spectrum emitted from one shell,
neglecting the emission from the other shells.
The steady spectrum can be interpreted as a superposition of those spectra.
The time (ks) labeling each thin line is for observers at Earth.
\label{fig:9}}
\end{figure}

The final results for our model with external photons are shown in
Figure \ref{fig:9}, where the model spectrum well agrees
with observed spectra from the radio to the TeV.
As we have explained in \S 2, the steady spectrum is
a superposition of emission from multiple shells at different $R$.
However, the steady photon spectrum is virtually determined
by the electron spectral shape at $R=3 R_0$,
because the rapid increase of the electron injection makes the electron density
reach a maximum at the end point of the injection/acceleration (see Figure \ref{fig:7}).
However, the emission spectrum from one shell for an observer
evolves as shown in Figure \ref{fig:9}.
The synchrotron component shows a hard-to-soft evolution.
This may be due to the decay of the magnetic field.
On the other hand, the SSC component, which has a peak around $10^{11}$ eV,
does not show a drastic evolution in its hardness.
The Klein--Nishina effect makes a peak at the energy that
is determined by the maximum energy of the electrons.
As a result, this peak energy is insensitive to the synchrotron peak energy.
The EIC emission, whose spectral peak is clearly seen in the single-shell spectra,
especially for the early period (10--32 ks), succeeds in reproducing the {\it Fermi} data.

While two-zone models are still promising, the success of the EIC model
encourages single emission-region models.
The EIC model needs another parameter set for the external photon field.
The essential parameters are its luminosity and peak photon energy,
because the details of the photon spectral shape are not so important.
Thus, the practical number of model parameters is 10
in this model (see the last part in \S \ref{sec:model}).
This number is still fewer than the DBP model in \citet{mrk421wide},
although the DBP model is not designed to address the radio spectrum.

\section{Variability in Mrk 421}
\label{sec:var}

The MAGIC telescope reported day-scale flux variations
and a clear correlation between TeV and X-ray fluxes of Mrk 421 \citep{mrk421magic}.
\citet{mrk421X-TeV} claimed
a possible lag ($\sim 2$ ks) of TeV flares relative to
soft X-ray flares, whereas TeV and hard X fluxes are well correlated
\citep[see also][]{mrk421opt-TeV}.

Spectral evolution obtained with {\it Suzaku} \citep{ush09}
indicates that the spectral peak energy shifts to a higher energy
with increasing flux in X-ray flares of Mrk 421.
Another interpretation \citet{ush09} claimed is that
two components, ``steady'' and ``variable,''
coexist in X-ray flares.
The ``variable'' component is described
by a broken power-law, while the ``steady'' component has
an exponential cutoff at $\sim 1$ keV.
In this section, based on the picture of the ``steady''
and ``variable'' components, we argue the spectral evolution
in flares in Mrk 421.
As shown in \citet{mrk421swift}, the flare spectra may provide
a signature of Fermi-II acceleration.
Note that we do not intend to fit individual flare spectra.\footnote{
The minimum variability timescale in observations
is also shorter than that we calculated here.
Our timescale, however, is within the distribution of the flare timescale.}
We just probe the qualitative behaviors of the flare spectra
with our time-dependent code.

In our steady flow approximation,
the identical shells are continuously ejected from $R=R_0$,
as shown in Figure \ref{fig:1}.
In order to simulate flares in Mrk 421,
we replace one shell in the sequence of the shells with a shell that
has a different parameter set from the other shells.
Then, the time-dependent contribution from the replaced shell
will produce a flare on the steady emission due to the other shells.
In this section, the model for the steady emission
is the same as the model with the external photons
in \S \ref{421EIC}.

\subsection{Variable plasma parameters}
\label{sec:varpp}

First, we propose a model in which the replaced shell
has a larger diffusion coefficient
and lower magnetic field than those for the other shells.
The other parameters are the same as those for the other shells
except for $\dot{N}'_{\rm e}$.
The magnetic field is taken to be $B_0=0.06$ G
and $K_0$ is $1.5$ times the value for the other shells.
To harden the electron/photon spectrum of the flare,
the injection rate is also changed to
$\dot{N}'_{\rm e}=\dot{N}_0 (R/R_0)^5$
(remember that $\dot{N}'_{\rm e} \propto R^7$ for the other shells),
where $\dot{N}_0=4.9 \times 10^{44}~\mbox{s}^{-1}$,
five times larger than the steady model.
Since the highest energy electrons are
dominated by those injected earlier, 
we expect that emission from high-energy electrons will
show a large change, while emission from low-energy electrons will be unaffected.

\begin{figure}[htb!]
\centering
\epsscale{1.0}
\plotone{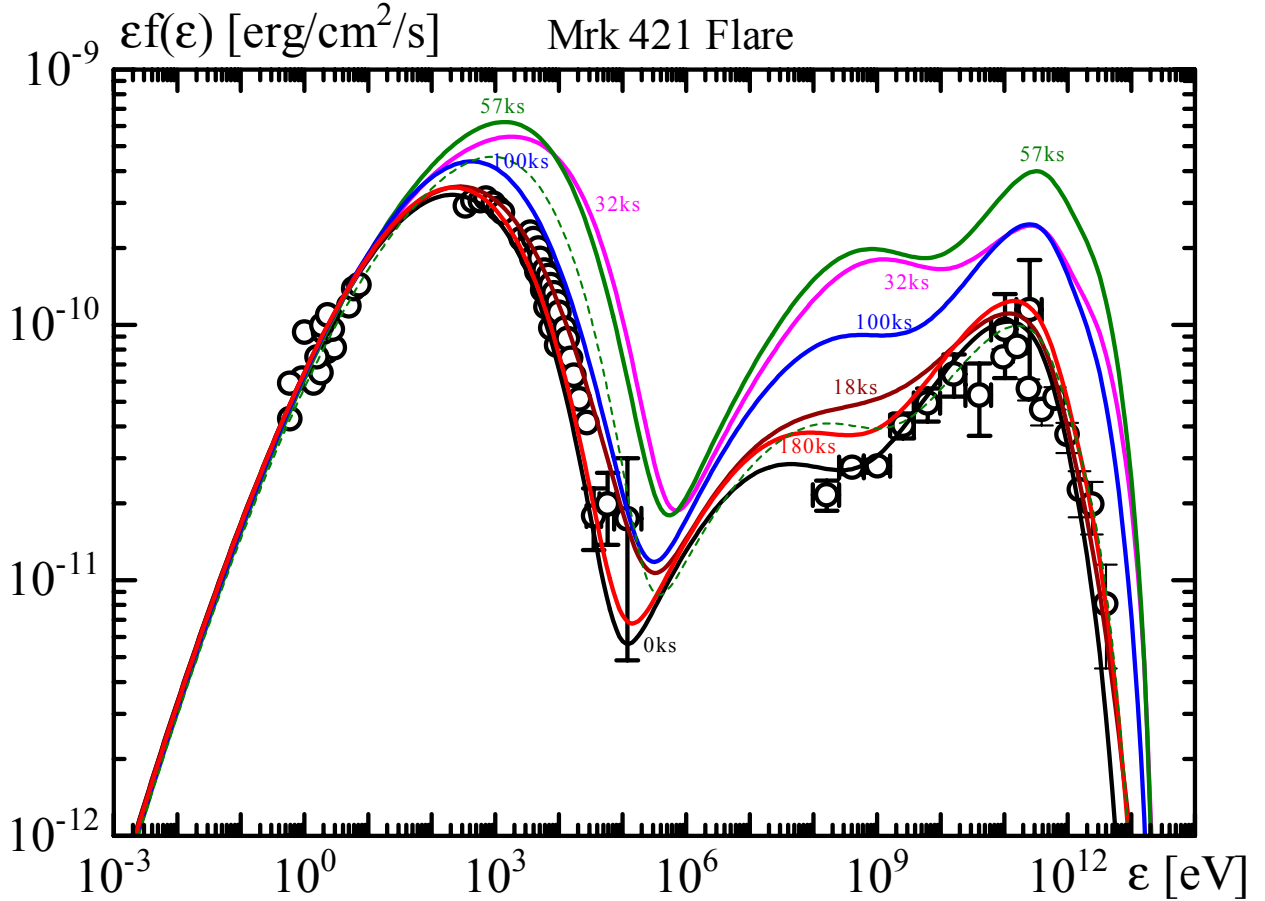}
\caption{Flare from a high $K'$ ($1.5$ times) and low magnetic field
($B_0=0.06$ G and $\dot{N}_0=4.9 \times 10^{44}~\mbox{s}^{-1}$) shell for Mrk 421
(see \S \ref{sec:varpp}).
The time (ks) for observers at Earth is denoted for each line.
The dashed line is the spectrum at 57 ks for the model
with the same magnetic field ($B_0=0.13$ G
and $\dot{N}_0=9.8 \times 10^{43}~\mbox{s}^{-1}$) as that in the other shells.
\label{fig:10}}
\end{figure}

Figure \ref{fig:10} shows the spectral evolution for this model.
The larger $K_0$ shifts the peak energy of the synchrotron component
to a higher energy.
This is similar to the observed hardening in hard X-ray bands.
This model yields a significant TeV flare as well.
When we do not change the magnetic field and $\dot{N}_0$
($K_0$ is changed, as explained),
a significant TeV flare does not appear (see the dashed line in Figure \ref{fig:10}).
This is because of the Klein--Nishina effect.
Even if the maximum energy of electrons is increased by the larger $K_0$,
the efficiency of the IC emission for such high-energy electrons
is very low.
Such electrons cool radiatively via nearly only synchrotron emission.
Therefore, to synchronize a TeV flare with an X-ray flare,
we need not only higher $K_0$ but also lower $B_0$,
which enhances the IC emission efficiency.
Since the external photon field is common for all the shells in this model,
a TeV flare inevitably accompanies a GeV flare, which is
more prominent than the TeV flare.
Short time flaring behavior in the GeV band for Mrk421
has not been detected until now, probably because BL Lac objects are
relatively weak GeV emitters.
Future studies will  be valuable to support or reject the EIC scenario of GeV emission.
%PKS1510-089

The required anti-correlation in $B_0$ and $K_0$ may seem awkward.
As shown in eq. (\ref{Dep}), $D \propto \varepsilon_{\rm e} k
|\delta B^2|_k/B \propto |\delta B^2|_k \propto
\delta B_0^2 B^{-q} \varepsilon_{\rm e}^q$, where
$\delta B_0^2$ is the normalization coefficient of $|\delta B^2|_k$.
Thus, if a decrease in $B$ does not accompany a change of $\delta B_0^2$,
the diffusion coefficient can be enhanced, because
the resonant wavenumber $k$ shifts lower as $B$ decreases.
However, if $\bar{\xi}$ is determined by the Alfv\'en velocity,
$K \propto B^2 |\delta B^2|_k \propto \delta B_0^2 B^{2-q}$.
Since $q<2$, an anti-correlation in $B$ and $\delta B_0^2$ is required
to enhance the diffusion coefficient.

\begin{figure}[htb!]
\centering
\epsscale{1.0}
\plotone{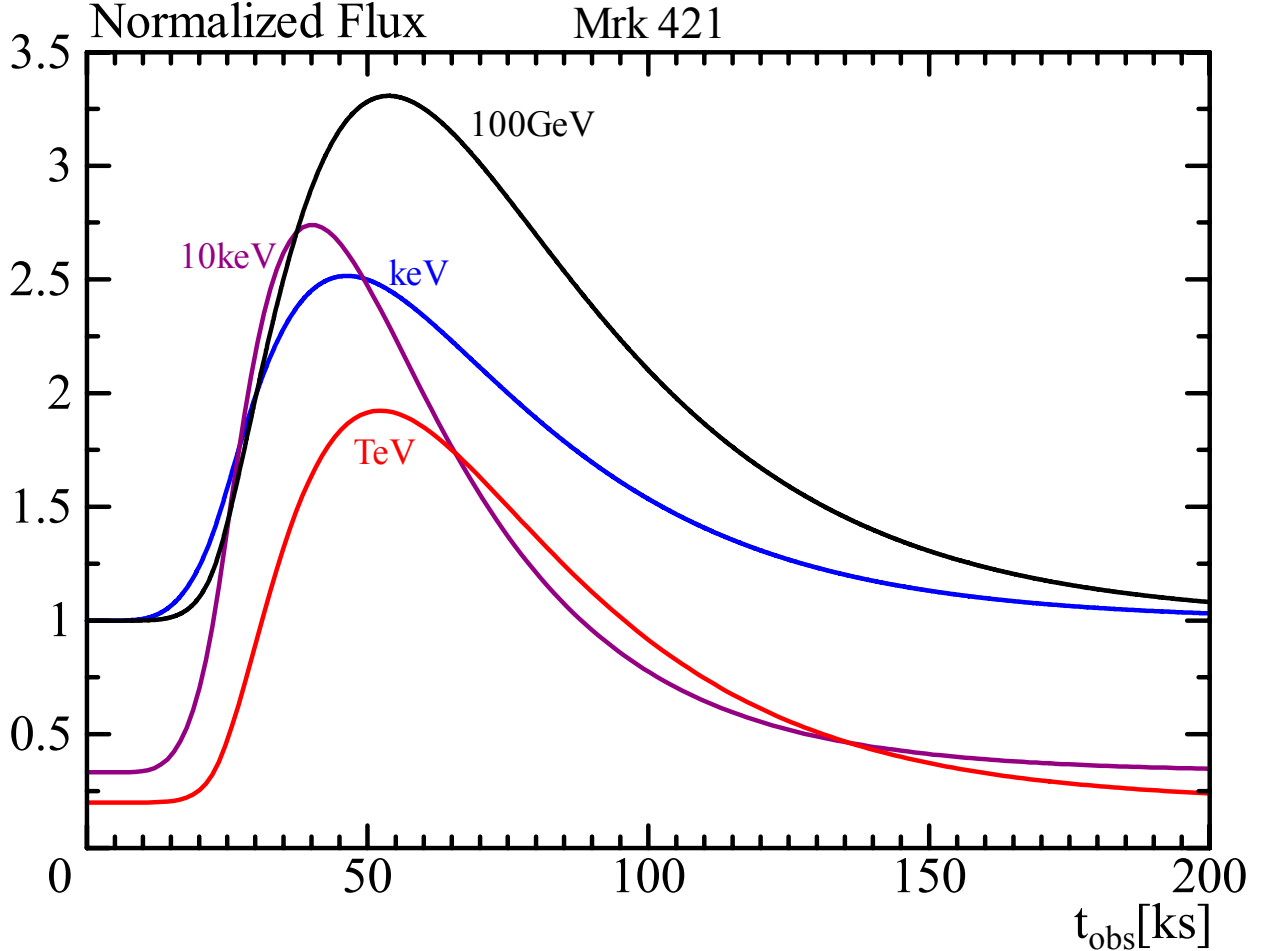}
\caption{Lightcurves for Fig. \ref{fig:10}.
\label{fig:11}}
\end{figure}

As we have discussed previously,
a flare in the hard X-ray band is emitted by the highest-energy electrons,
while the origin of the TeV flare is SSC emission from relatively lower-energy electrons.
Because radiative cooling is very effective for the highest-energy electrons,
the hard X-ray flare ceases faster than flares in other energy bands,
as shown in Figure \ref{fig:11}.
This tendency is not consistent with the observed synchronicity of hard X-ray
and TeV flares or the hard lag between X-ray bands.
The time-dependent simulations by \citet{time-dep}
also failed to reproduce this observed feature.
The slight lag of 100 GeV--TeV lightcurves relative to the soft X-ray flare
seem to be reproduced by our simulations.
The model lightcurves show long tails, while typical lightcurves from blazars
are almost symmetric in their rise and decay shape.
This long tail is not due to the curvature effect, namely
the contribution of off-axis emission.
Since radiative cooling is inefficient for most of the electrons,
the decay of the flares is regulated by adiabatic cooling.
Unless sudden shutdown of the emission is artificially adopted,
emission from slowly cooling electrons yields long tails in their lightcurves.

The above two problems, the early termination of the hard X-ray flare
and asymmetric lightcurves, are inevitable in our model.
We have replaced only one shell, changing the physical parameters
to produce a flare.
This implies that a partial and discrete transition of the physical parameters
occurs in the outflow.
Realistic outflows may have a gradual parameter change in a wider spatial range.
The symmetric lightcurve may be a result of this gradual parameter change.
Moreover, if the onset of the magnetic field decay is faster than the
increase of the diffusion coefficient,
the observed delay of hard X-ray flares should be reproduced.
Thus, the hard X-ray delay requires different evolution
of the magnetic field and electron injection.

\subsection{Variable Lorentz factor}
\label{sec:varLF}

A fluctuation of the bulk Lorentz factor may cause a flare as well.
Shifts of the spectral peak energies are naturally expected for
a photon source with a higher $\Gamma$.
Strictly speaking, we cannot embed a faster shell in a steady flow of a constant $\Gamma$.
Such a shell interacts with the precedent shell, and may
be decelerated by shocks.
Actual outflows may not be completely continuous.
Hence, postulating a quasi-steady outflow as a background,
we simply add the contribution of the faster shell
to the emission discussed in \S \ref{sec:Mrk} here.

Given a synchrotron luminosity,
a higher $\Gamma$ leads to a lower synchrotron photon density
in the shell frame.
In order to produce simultaneous X-ray and TeV flares,
a weaker magnetic field is required even in this case.
This means that a larger $K'$ is also required to shift the spectral peaks higher.
Here, a shell with $\Gamma=30$ is assumed to be the origin of the flare.
We adopt the same $R_0$ as before, but the high $\Gamma$ leads
to a narrower width $W'=5 \times 10^{15}$ cm.
Other parameters are $\Delta T'_{\rm inj}=2 W'/c$,
$B_0=0.03$ G, $K'=K_0 (R/R_0)^{-1}$ with
$K_0=3.9 \times 10^{-2}~\mbox{eV}^{1/3}~\mbox{s}^{-1}$, and
$\dot{N}'_{\rm e}=\dot{N}_0 (R/R_0)^5$ with
$\dot{N}_0=4.9 \times 10^{44}~\mbox{s}^{-1}$.

\begin{figure}[htb!]
\centering
\epsscale{1.0}
\plotone{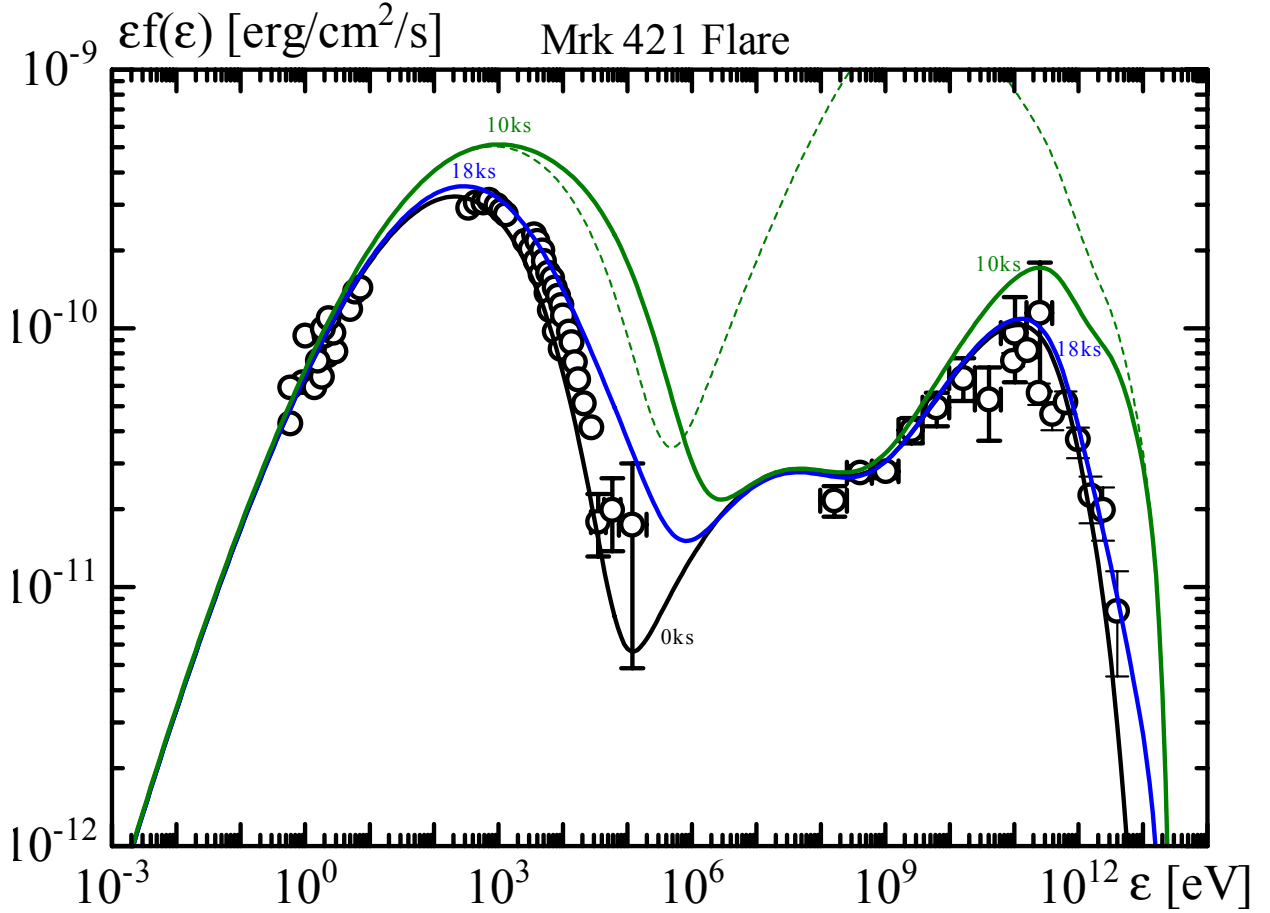}
\caption{Flare from a high-$\Gamma$ ($\Gamma=30$) shell for Mrk 421
(see \S \ref{sec:varLF}).
The time (ks) for observers at Earth is denoted for each line.
The solid lines are model spectra neglecting the EIC emission.
The dashed line is the spectrum at 10 ks for the model
including the EIC effect.
\label{fig:12}}
\end{figure}

The obtained spectra are plotted in Figure \ref{fig:12}.
If we neglect the EIC emission in this model,
flares are seen in only X-ray and TeV energy bands (solid lines).
However, a single outflow model with the external photons
imposes the EIC emission on the flare source.
The higher $\Gamma$ enhances the efficiency of the EIC;
given the electron total number and energy distribution
in the shell frame, the EIC luminosity is proportional
to $\Gamma^6$ (a Doppler factor $\delta \sim \Gamma$ is assumed),
while the synchrotron luminosity is $\propto \Gamma^4$
\citep[see e.g.,][]{der02}.
Hence, the amplification of the GeV flare due to the EIC emission
is very large (the dashed line in Figure \ref{fig:12}).
If this huge GeV flare is not observationally favorable,
the high-$\Gamma$ model with the external photons will be rejected.
In this case, $\Gamma$ should be almost constant,
or a different source for the GeV steady emission (no external radio source)
may be required.

When shells with different values of $\Gamma$ are injected,
they collide and particles are accelerated by first- and second-order
Fermi processes. \citet{bot10} elaborated
on the emission properties from such collisions and showed that the 
various types of
the evolution of the emission spectrum are induced by such collisions.
We have neglected such effects in the EIC models, which
may be observationally constrained based on the high sensitivity to $\Gamma$.
This should be tested in future studies.

\begin{figure}[htb!]
\centering
\epsscale{1.0}
\plotone{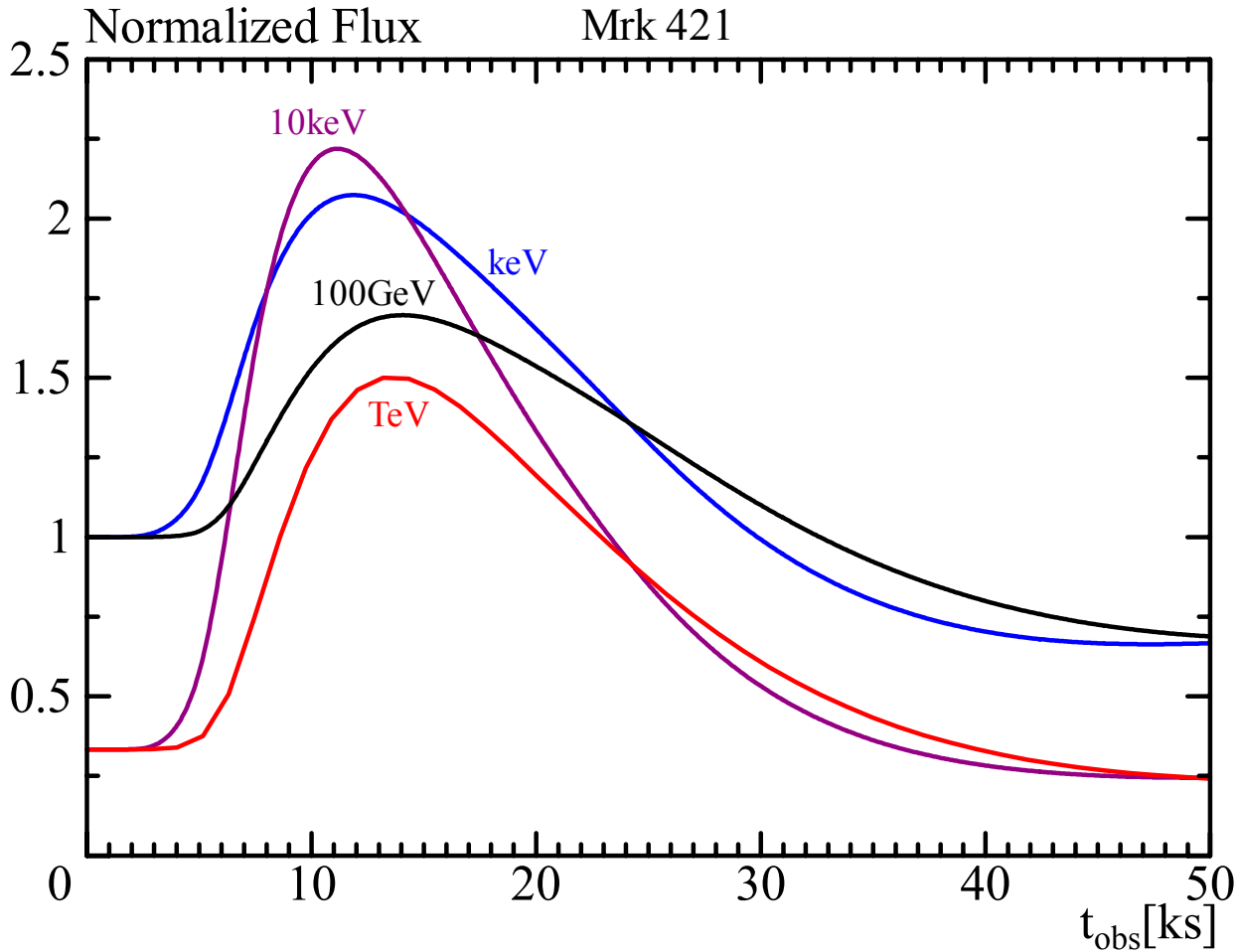}
\caption{Lightcurves for Fig. \ref{fig:12}.
\label{fig:13}}
\end{figure}

The lightcurves for the model without the EIC emission
are plotted in Figure \ref{fig:13}.
The high $\Gamma$ leads to a short variability timescale ($\propto \Gamma^{-2}$)
compared with the model in \S \ref{sec:varpp}.
The qualitative behavior is similar to the case in Figure \ref{fig:11}.
The very weak magnetic field extends the cooling time scale for
the highest-energy electrons.
Thus, the early termination of the hard X-ray flare is not prominent
compared with the model in \S \ref{sec:varpp}.

\subsection{Shock acceleration}
\label{sec:varsh}

While the quasi-steady emission may be due to the Fermi-II
acceleration, the flare phenomena may be attributed to
shocks in the outflow.
The interpretation in \citet{ush09} is compatible with such a picture.
As a model with a combination of Fermi-I and Fermi-II processes,
\citet{wei10,wei10b} calculated electron and photon spectra,
dividing the  blazar region into acceleration and radiation zones.
The accelerated electrons escape from the acceleration zone and
are injected into the radiation zone.  By changing the particle injection,
they obtained the light curves for 1ES 1218+30.4 and PKS 2155-034.

\begin{figure}[htb!]
\centering
\epsscale{1.0}
\plotone{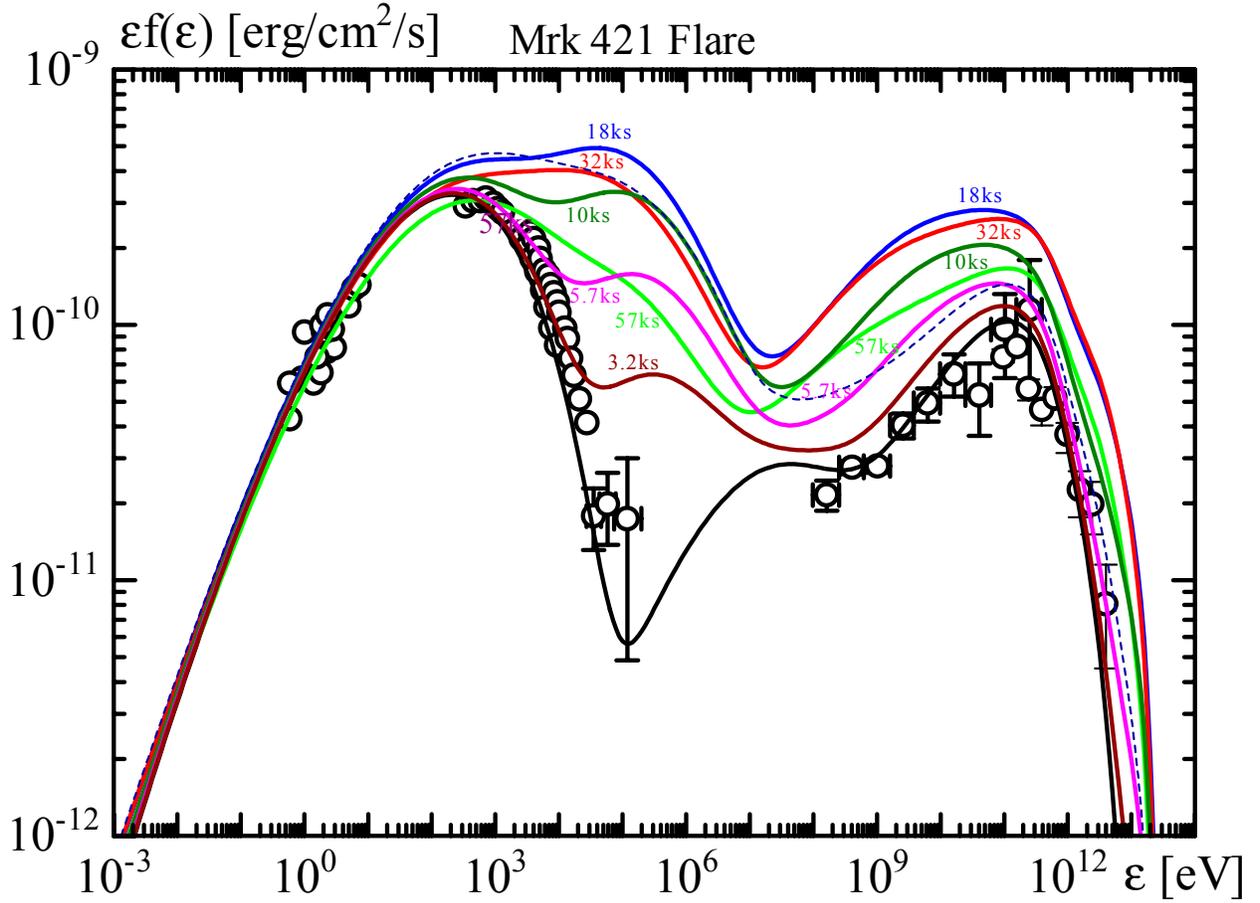}
\caption{Flare from a shocked shell for Mrk 421 (see \S \ref{sec:varsh}).
The time (ks) for observers at Earth is denoted for each line.
The solid lines are model spectra
with $B_0=0.06$ G and $E_{\rm e,iso}=5 \times 10^{51}$ erg.
The dashed line is the spectrum at 18 ks for the model
with $B_0=0.13$ G and $E_{\rm e,iso}=2 \times 10^{51}$ erg.
\label{fig:14}}
\end{figure}

Within our picture, we also test the Fermi-I model with our code.
We inject shock-accelerated electrons of the single power-law
with an exponential cutoff into the flaring shell.
The power-law index is $p=2$
and the cutoff Lorentz factor is $\gamma_{\rm e,max}=10^7$.
The minimum Lorentz factor is taken to be $\gamma_{\rm e,min}=15$.
The bulk Lorentz factor is $\Gamma=15$ and
the shell width is $W'=1.0 \times 10^{16}$ cm,
the same as those in the steady component.
The injection is assumed to be constant
over a time scale $\Delta T'_{\rm inj}=W'/c$
and we neglect the reacceleration by turbulence.
The total energy of electrons is $E_{\rm e,iso}=5 \times 10^{51}$ erg
in spherically symmetric evaluation ($E_{\rm e}=E_{\rm e,iso} \theta_{\rm j}^2/2
=1.1 \times 10^{49}$ erg).
Even in this model, a weak magnetic field is required
($B_0=0.06$ G) to produce a TeV flare
(see the dashed line in Figure \ref{fig:14} for the model
with $B_0=0.13$ G and $E_{\rm e,iso}=2 \times 10^{51}$ erg).

\begin{figure}[htb!]
\centering
\epsscale{1.0}
\plotone{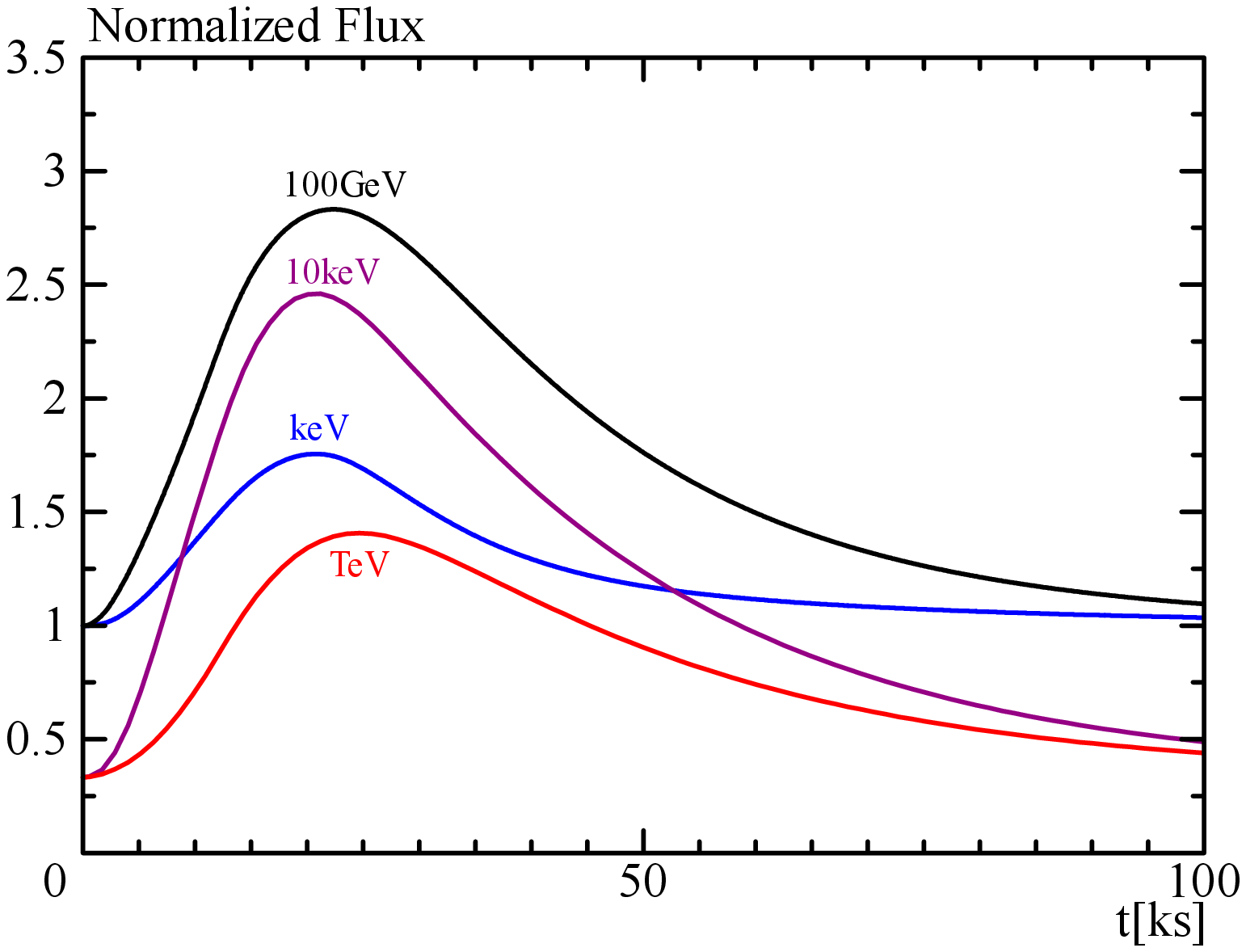}
\caption{Lightcurve for Fig. \ref{fig:14}.
\label{fig:15}}
\end{figure}

As shown in Figure \ref{fig:14}, the synchrotron spectra
show flat shapes (the photon index is $\sim 2$) in the X-ray band.
These values are significantly different from the other models.
The lightcurves in Figure \ref{fig:15} show
coincident peaks at $\sim 20$ ks from keV to 100 GeV.
This is due to the continuous injection of the high-energy electrons.
The electron injection and cooling balance each other in the high-energy regions,
so the electron energy distribution remains quasi-steady until the electron
injection stops.
The termination times of the emissions are controlled
by the electron injection.
The slight delay of the TeV lightcurve may come from the evolution
of the seed X-ray photons.

\section{Summary and Discussion}
\label{sec:sum}

In this paper, we have simulated the temporal evolution
of high-energy electrons and photon production
in relativistically outflowing shells.
Our numerical code can follow the electron distribution
with the effects of
the electron injection, acceleration, synchrotron cooling,
and IC cooling.
The full non-linearities
of IC cooling including the Klein--Nishina effect, are taken into account.
We have considered the Fermi-II process as the electron acceleration
mechanism, while there are other candidates for the acceleration mechanism,
such as Fermi-I.
The Fermi-II process, driven by some kind of turbulence
in the outflows, can naturally make electron spectra
harder than those predicted by the
simplest version of diffusive shock acceleration theory.
As opposed to the shock acceleration in supernova remnants,
the maximum energy of the electrons is expected to be far below
that in the Bohm limit.
Those characteristics are favorable to explain blazar photon spectra.
In this method and model,
the diversity in the temporal evolution
of the electron injection and acceleration
can be expected to generate a variety of photon spectral shapes.

We have modeled steady photon emission by superposition
of time-evolving emission from continuously ejected multiple shells.
The photon spectrum of the TeV blazar 1ES 1101-232
is well reproduced by a simple model with a constant
injection rate and diffusion coefficient.
For Mrk 421, which shows a softer spectrum than that
in 1ES 1101-232, we need to adjust the evolution
of the electron injection rate etc. to fit the spectrum.
A power-law evolution of $\dot{N}'_{\rm e} \propto R^7$
makes a curved electron spectrum, which produces a good fit
to the observed synchrotron spectrum from the radio to X-ray bands.
An advantage in our model is that we do not need to introduce
unprescribed energy scales as break energies in the electron spectrum.
However, the required rapid growth of the injection rate has not been
theoretically justified yet.
Future progress in the study of the injection processes with time-dependent ways
will be important to examine the validity of the model.

Our Fermi-II model explains the radio data as well as the optical and X-ray data
as the emission from a single source.
In most of preceding models based on shock acceleration, which fit the
optical and X-ray data of blazars, the electron density in the low-energy
range is much higher than in our model (see Figure \ref{fig:7}), so that the
synchrotron self-absorption effect is much stronger. As a result,
the radio data have difficulty explaining simultaneously the
optical and X-ray data by one-zone shock acceleration models.
These data are frequently explained by the superposition of multi-zone
self-absorbed emission \citep[e.g.][]{kon81}.
The picture we proposed is different from such models.
%This multi-zone
%model appears to be consistent with the observed radio core shifts
%towards the central engine with increasing frequency
%(e.g., Hada et al. 2012,
%O'Sullivan \& Gabuzda, 2009, MNRAS, 400, 26).
%On the other hand, If our model is correct for blazars for
%which
%the radio core shifts are observed, some additional mechanisms should be
%introduced, e.g., the radial dependence of the minimum electron energy.

In this paper, we have conservatively assumed the Kolmogorov type
of turbulence, $q=5/3$. An alternative way to make a soft electron spectrum
is to adopt a larger value of $q$.
Even if $\dot{N}'_{\rm e}$ and $K'$ are constant,
the hard-sphere scattering \citep[$q=2$, see e.g.][]{par95}
leads to a soft electron spectrum as shown in Figure \ref{fig:2}.
This case implies that the acceleration timescale is independent
of the particle energy, which is similar to the original idea
of \citet{fer49}.
Another possibility is the effect of particle escape \citep[e.g.][]{bec06};
we have not included this effect.
Especially for the model with $q=2$,
the escape timescale is independent of the particle energy
so that the effect can be important.
The escape effect will not only change the
spectral index of the electron distribution,
but in some cases may also introduce cutoffs in $N_{\rm e}$.
We may need at least two zones, an acceleration region
and an emission region, to simulate such models without neglecting
the contribution of the escaped particles.
Note that the model of \citet{wei10,wei10b} is a two-zone model.
However, the accelerated electrons are injected
in the emission zone uniformly and the effect of geometrical 
separation of the acceleration and emission zones has not been considered.

It is interesting that the obtained electron spectrum is close to the log-parabolic
function in the most important energy range.
In \citet{mas04}, the origin of this shape is attributed to
the energy dependence of the escape probability.
However, our time-dependent calculations have not included the escape effect.
The analytical study of the Fermi-II process by \citet{par95}
based on Green's functions may be a meaningful hint for
this spectral shape.
Since most high-energy electrons are
injected at early times, their spectral shape is primarily determined by the Green's
function for a single injected energy.
The spectral shapes around the synchrotron peak predicted by
the DBP and log-parabolic models are hard to distinguish
from our model.
Thus, to search for the signature of the electron minimum energy
required in those analytical models,
future infrared and submillimeter observations will be required.

In order to reproduce the GeV flux for Mrk 421
by our single emission-region model,
an external radio photon field is needed.
The radio photons interacting with high-energy electrons in the outflow
can be up-scattered to GeV energies.
The required radio luminosity $4.9 \times 10^{38}~\mbox{erg}~\mbox{s}^{-1}$
is far below the bolometric luminosity $1.4 \times 10^{43}~\mbox{erg}~\mbox{s}^{-1}$
(assuming $\theta_{\rm j}=1/15$).
Alternatively, an additional emission region may contribute as
a GeV photon source.
Correlation analyses of flux variabilities between GeV
and another band may provide a clue to the GeV emission region.
While significant variability (a factor of about three)
in the flux has been reported \citep{mrk421wide},
the correlations with X-ray or TeV variabilities seem still ambiguous
for determining the model.

By replacing a shell in the sequence of the identical shells
and changing the parameters,
we simulate flare phenomena.
In this method, the flare lightcurves show asymmetric shapes.
The flare emission gradually fades out via adiabatic cooling.
To reproduce the symmetric lightcurves as is frequently seen in
blazar flares, gradual changes of the parameters may be required,
while our models correspond to discrete changes of the parameters.
The cooling time of the electrons that emit hard X-rays
is quite short.
Therefore, a gradual cessation of the electron injection or acceleration
in this highest energy range may be required to synchronize
the peak times of the hard X-ray and TeV lightcurves.
A sudden shut down of acceleration/injection would lead to
an early hard X-ray termination.
The most critical aspect to produce simultaneous flares
in the X-ray and TeV bands is to decrease the magnetic field.
The Klein--Nishina effect prevents TeV flares caused by
a growth of the diffusion coefficient that increases
the electron maximum energy.
An enhancement of the SSC emission efficiency by weakening the magnetic field
is required to generate a TeV flare.
The required anti-correlation between the fluxes and the magnetic field
is a challenging problem.

For the EIC model, high variability of the bulk  Lorentz factor $\Gamma$
is not favorable. Since the EIC emission is sensitive to $\Gamma$,
the observed GeV variability
strictly constrains the fluctuation of $\Gamma$ by about a factor of three.

\begin{acknowledgments}
We appreciate the anonymous referee for valuable comments
that improved our paper significantly.
This study is partially supported by Grants-in-Aid for Scientific Research
No.25400227 and 24540258 from the Ministry of Education,
Culture, Sports, Science and Technology (MEXT) of Japan (KA),
and JSPS Research Fellowships for Young Scientists No.231446 (KT).
\end{acknowledgments}

\end{document}